\documentclass{aa}

\usepackage{graphicx}
\usepackage{txfonts}
\usepackage{hyperref}
\usepackage{color}
\usepackage{comment}
\usepackage{amsmath,bm}
\usepackage{siunitx}
\usepackage{dcolumn}

\begin{document}

   \title{The middle-aged  
   pulsar PSR J1741--2054 and its bow-shock nebula in 
   far ultraviolet\thanks{Based on observations made with 
the NASA/ESA Hubble Space Telescope, obtained at the Space Telescope Science Institute, which is operated by the Association 
of Universities for Research in Astronomy, Inc., under NASA contract NAS 5-26555. These observations are associated with program \#17155.}
}

   \subtitle{}

   \author{Vadim Abramkin
          \inst{1}
          \and
          George G. Pavlov
         \inst{2}
          \and
          Yuriy Shibanov
          \inst{1}
          \and
          B. Posselt
          \inst{3,2}
          \and
          Oleg Kargaltsev
          \inst{4}
          }

   \institute{Ioffe Institute, Politekhnicheskaya 26, St. Petersburg, 195251, Russia
              \\
              \email{vadab2077@gmail.com}
         \and
         Pennsylvania State University, Department of Astronomy \& Astrophysics, 525 Dave Lab., University Park, PA 16802, USA
             \\
             \email{ggp1@psu.edu}
        \and
        Oxford Astrophysics, University of Oxford, Denys Wilkinson Building, Keble Road, Oxford OX1 3RH, UK
        \and
        The George Washington University, Department of Physics, 725 21st Street NW, Washington, DC 20052, USA
             }


 
  \abstract
   {Far-ultraviolet (FUV) observations of 
   pulsars allow one to measure surface temperatures of neutron stars and study their thermal evolution. 
   Some pulsars can exhibit 
   FUV bow-shock nebulae (BSNe), providing an additional tool to probe  
   the interstellar medium 
   and study the pulsar's properties. 
   The nearby middle-aged gamma-ray 
   pulsar J1741--2054 and its pulsar wind nebula (PWN) have been studied in X-rays, and its 
   BSN has been investigated in the Balmer lines,
   but they have never been observed in FUV.
   }
   {To further study the thermal and magnetospheric emission from PSR J1741--2054 and the BSN properties, we observed 
   them in the FUV range with the Hubble Space Telescope (HST). }
   {We imaged the target in two FUV filters of the
   HST's ACS/SBC detector. 
   We also re-analyzed previous optical observations of the pulsar and its 
   BSN.
   We fit the pulsar's FUV-optical spectrum separately and together with its X-ray spectrum.
   }
   {We found that 
   the pulsar's FUV-optical spectrum consists of a thermal and nonthermal components. A joint fit of the FUV-optical and X-ray spectra with combinations of a nonthermal 
   and thermal 
   components showed a hard optical nonthermal spectrum with a photon index $\Gamma_{\rm opt} \approx 1.0$--1.2 and a softer X-ray component, $\Gamma_X \approx 2.6$--2.7. The thermal emission is dominated by the cold component with the temperature $kT_{\rm cold}\approx 40$--50 eV and emitting sphere radius $R_{\rm cold}\approx 8$--15 km, at $d=270$ pc. An additional hot thermal component, with $kT_{\rm hot}\sim 80$ eV and $R_{\rm hot}\sim 1$ km, is also possible. Such a spectrum resembles the spectra of other middle-aged pulsars, but it shows a harder (softer) optical (X-ray) nonthermal spectrum.
   We detected the FUV BSN, the first one associated with a middle-aged pulsar. Its closed-shell morphology is 
   similar to the H$\alpha$ BSN morphology, while its 
   FUV flux, $\sim10^{-13}$ erg cm$^{-2}$ s$^{-1}$, is a factor of $\sim 4$ 
   higher than the H$\alpha$ flux. 
   This FUV BSN 
   has a higher surface brightness than the two previously known ones. 
   }
   {}


   \maketitle
%

\section{Introduction}
\label{intro} 

Rotation powered pulsars emit nonthermal radiation, generated by relativistic
particles in the pulsar magnetosphere and observable from radio through $\gamma$-rays. 
In addition to this magnetospheric radiation, 
thermal emission from the neutron star (NS) surface can be observed in a narrower wavelength range, from the optical through soft X-rays.
The thermal emission is buried under the powerful magnetospheric emission in very young pulsars, but at pulsar ages of  $\sim 100$ kyr the magnetospheric emission becomes fainter, and the 
thermal emission emerges, with a maximum flux in the extreme UV or soft X-rays.
This thermal emission may include a cold component associated with the residual heat of the cooling NS,
and a hot component from a fraction of the NS surface,
possibly heated by relativistic particles precipitating from the pulsar magnetosphere or by an anisotropic heat transport from the NS interior. 
When the pulsar becomes older than $\sim 1$ Myr, 
the NS cooling results in 
the cold component being only detectable in UV-optical, but the hot thermal component can still be seen in soft X-rays, together with the (faint) magnetospheric component. Thus,
multiwavelength observations of middle-aged ($\tau \sim 0.1$--1\,Myr) pulsars allow one to study all the components of the NS's electromagnetic radiation, which is 
particularly important for understanding 
NS physics and evolution (e.g., \citealt{Pavlov2002}). In particular, observations of middle-aged pulsars in the optical-UV range allow one to separate the nonthermal component, which usually dominates in the optical part of the spectrum, and compare it with the nonthermal X-ray component, as well as detect the Rayleigh-Jeans tail of the cold thermal component and measure the temperature of the bulk of the NS surface. However, 
because of the intrinsic faintness of the optical-UV emission from 
middle-aged pulsars, only three of them (PSR\,B0656+14, Geminga, and PSR\,B1055--52)  have been well studied in this range (see, e.g., \citealt{Posselt2023}, and references therein). Therefore, adding just one object to this small sample 
is important for understanding the NS properties and NS evolution. In this paper, we report on a recent far-UV (FUV) detection of one more middle-aged pulsar with the Hubble Space Telescope (HST) and compare its optical-UV and X-ray properties with those of the other detected middle-aged pulsars.

PSR J1741--2054 (J1741 hereafter) was first detected as a $\gamma$-ray source 3EG J1741--2050 with the Energetic Gamma-Ray Experiment Telescope aboard the Compton Gamma-ray Observatory 
\citep{Hartman1999}. Gamma-ray pulsations with a period $P=413.7$ ms and a period derivative $\dot{P}=1.698\times 10^{-14}$\,s\,s$^{-1}$ 
were discovered with the {\em Fermi} Large Area Telescope (LAT) 
\citep{Abdo2009}. These $P$ and $\dot{P}$ values correspond to spindown power (energy loss rate) $\dot{E}=9.5\times 10^{33}$\,erg\,s$^{-1}$, characteristic spindown age $\tau=386$ kyr, and surface magnetic field $B=2.7\times 10^{12}$ G. Radio pulsations from this pulsar with one pulse per period were discovered by \citet{Camilo2009}. The small dispersion measure ${\rm DM} = 4.7$\,pc\,cm$^{-3}$ 
at the Galactic coordinates $l=6\fdg42$, $b=4\fdg91$ corresponds to the distance $d=270$\,pc    
in the Galactic electron distribution model by \citet{Yao2017}, or $d=380$\,pc
 in the model by \citet{Cordes2002}. The 0.1--300 GeV energy flux $F_\gamma \approx 1.4\times 10^{-10}$\,erg\,cm$^{-2}$\,s$^{-1}$ corresponds to the (isotropic) luminosity 
 $L_\gamma \equiv 4\pi d^2 F_\gamma = 1.2\times 10^{33}d_{270}^2$\,erg\,s$^{-1}$, where $d_{270} = d/270\,{\rm pc}$.  \citet{Camilo2009} also found that the $\gamma$-ray pulse profile occupies
1/3 of pulsation period and has three closely spaced peaks with the first peak lagging the radio pulse by $0.29 P$ (see also \citealt{Ray2011}; \citealt{Smith2023}). 

The X-ray properties of the pulsar were studied with {\em Swift} \citep{Camilo2009}, {\em Chandra} \citep{Auchettl2015,Karpova2014,Romani2010} and {\em XMM-Newton} \citep{Marelli2014}.
The phase-integrated X-ray spectrum of J1741 can be described as a combination of thermal and nonthermal components, absorbed by the interstellar medium (ISM) with $N_H \simeq 1.2\times 10^{21}$\,cm$^{-2}$. The thermal component, presumably emitted from the 
NS surface, is characterized by a blackbody (BB) temperature $kT\simeq 60$ eV and equivalent sphere radius $R\simeq 3.5 d_{270}$ km, 
while the nonthermal (presumably magnetospheric) spectrum fits a power law (PL) with a photon index $\Gamma\simeq 2.7$. 
The 0.3--10 keV unabsorbed fluxes of the nonthermal and thermal components are about $5.5 \times 10^{-13}$ and $7.6  \times 10^{-13}$\,erg\,cm$^{-2}$\,s$^{-1}$, respectively.
According to \citet{Marelli2014}, the X-ray pulsations show one broad pulse per period, with a lag of about $0.6P$ with respect to the radio pulse, and a pulsed fraction of 
35\%--40\% in both the 
thermal and nonthermal components.
{\em Chandra} observations allowed \citet{Auchettl2015} to measure the pulsar's proper motion:
$\mu_\alpha = -63\pm 12$\,mas\,yr$^{-1}$, $\mu_\delta = - 89\pm 9$ mas\,yr$^{-1}$. 
This corresponds to a transverse velocity 
$v_\perp = (140\pm 13) d_{270}$ km\,s$^{-1}$.

The X-ray images revealed a 
pulsar wind nebula (PWN) stretched in the 
northeast direction, which consists of a compact part of $\sim10''$ 
length and a diffuse ``trail'' of $\sim 100''$ (
$\sim 0.13 d_{270}$\,pc) length,  with a nonuniform brightness distribution. The 0.3--5 keV unabsorbed flux of the PWN is about $1.8\times 10^{-13}$\,erg\,cm$^{-2}$\,s$^{-1}$. 
The compact X-ray PWN component is enveloped by a horseshoe-like H$\alpha$ nebula, 
with a flux $F_{{\rm H}\alpha}=1.5\times 10^{-14}$ erg cm$^{-2}$ s$^{-1}$ and an apex at 1\farcs5 ahead of the pulsar
\citep{Romani2010,Brownsberger2014,Auchettl2015}.

Observations with the Very Large Telescope (VLT) in the $b_{\rm high}$ ($\lambda =4400$\,\AA, $\Delta\lambda = 1035$\,\AA) and $v_{\rm high}$ ($\lambda= 5570$\,\AA, $\Delta\lambda = 1235$\,\AA) filters allowed \citet{Mignani2016} to identify two pulsar counterpart candidates, consistent with the {\em Chandra} position of the pulsar. The fainter of the two objects, with  magnitudes $m_v=25.32\pm0.08$ and $m_b=26.45\pm 0.10$, was proposed as a more plausible candidate. The corresponding flux densities are well above the extrapolation of the X-ray thermal component to optical wavelength, but they are about 3 orders of magnitude below the PL component extrapolation. An arc-like structure southwest of the counterpart candidates, seen in the $b_{\rm high}$ filter only,
was identified as the H$\beta$/H$\gamma$ bow-shock nebula,
previously detected by \citet{Romani2010} in H$\alpha$. 

To confirm the candidate and understand the nature of its optical emission and its connection with the X-ray emission, additional deep observations of the field are required. Fortunately, J1741 was included in the 
HST program \#17155 ``The Legacy UV survey of 28 pulsars'' \citep{Kargaltsev2022}, aimed at constraining FUV fluxes and NS surface temperatures of 28 nearby pulsars. Despite the shortness of these observations (one HST orbit per pulsar), a number of targets, including J1741, have been detected. In this paper we present the results on the J1741 pulsar and its bow shock nebula (BSN hereafter) and discuss their connection with the X-ray and gamma-ray data.

\section{Observations}

\begin{figure}[t]
\center{\includegraphics[width=0.9\linewidth]{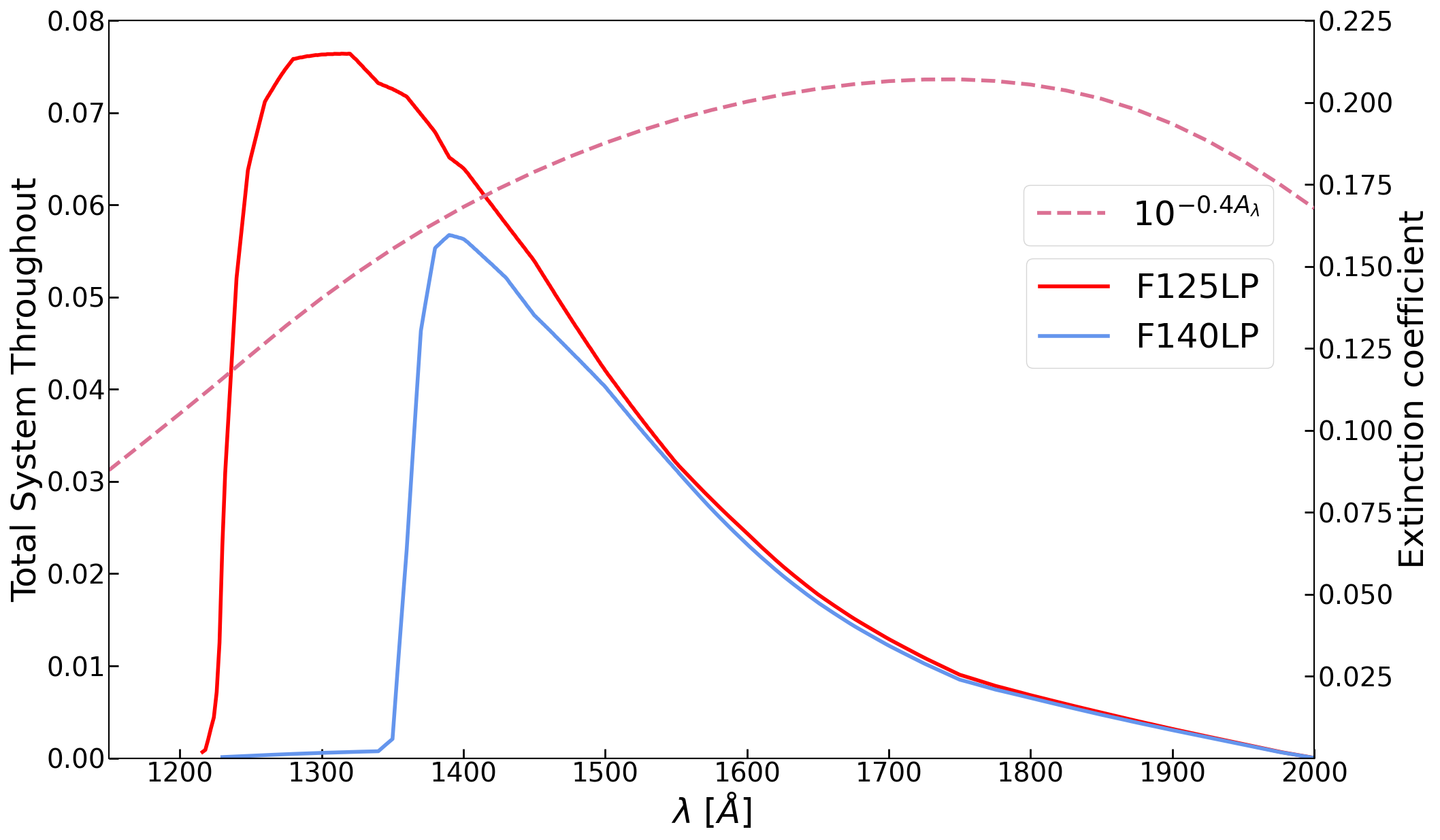}}
\caption{Wavelength dependencies of the throughputs $T(\lambda)$ of the ACS/SBC F125LP and F140LP filters (solid curves, values on the left) and 
the extinction coefficient $10^{-0.4 A_\lambda}$ at $E(B - V)= 0.22$ (the dashed curve, values on the right), most probable for J1741. 
\label{fig:filters}}
\end{figure}

We observed J1741 on 2023 June 12 (MJD\,60107; one HST orbit) with the Solar Blind Channel (SBC) of the Advanced Camera for Surveys (ACS) with a nominal field of view (FoV) of $\approx 31''\times 35''$
and the original pixel scale of about 
$0\farcs030\times 0\farcs034$
($0\farcs025\times 0\farcs025$ after drizzling). The SBC detector is a 
CsI microchannel plate with Multi-Anode Microchannel Array (MAMA) readout, sensitive from  1150\,\AA\   to 
2000\,\AA. The ACS/SBC operates in  
the ACCUM mode, producing a time-integrated image. 
In the shadow part of the orbit (1830 s) we employed the most sensitive F125LP filter, 
which cuts off the geocoronal Ly$\alpha$ emission. In the 
non-shadow part of the orbit 
(566 s) we used the F140LP filter, 
which additionally blocks Oxygen geocoronal lines. 
The wavelength dependencies of the full throughputs in this filters are shown in Figure~\ref{fig:filters}. The pipeline processed data were downloaded from the Barbara A.\ Mikulski Archive for Space Telescopes (MAST). Pipeline processing included calibration through the CALACS image processing package and correction for geometric distortion with the AstroDrizzle software from the {\tt DrizzlePac} package\footnote{\url{https://www.stsci.edu/scientific-community/software/drizzlepac}}. 

\begin{figure*}[t]
\begin{minipage}[h]{0.5\linewidth}
\center{\includegraphics[width=0.85\linewidth]{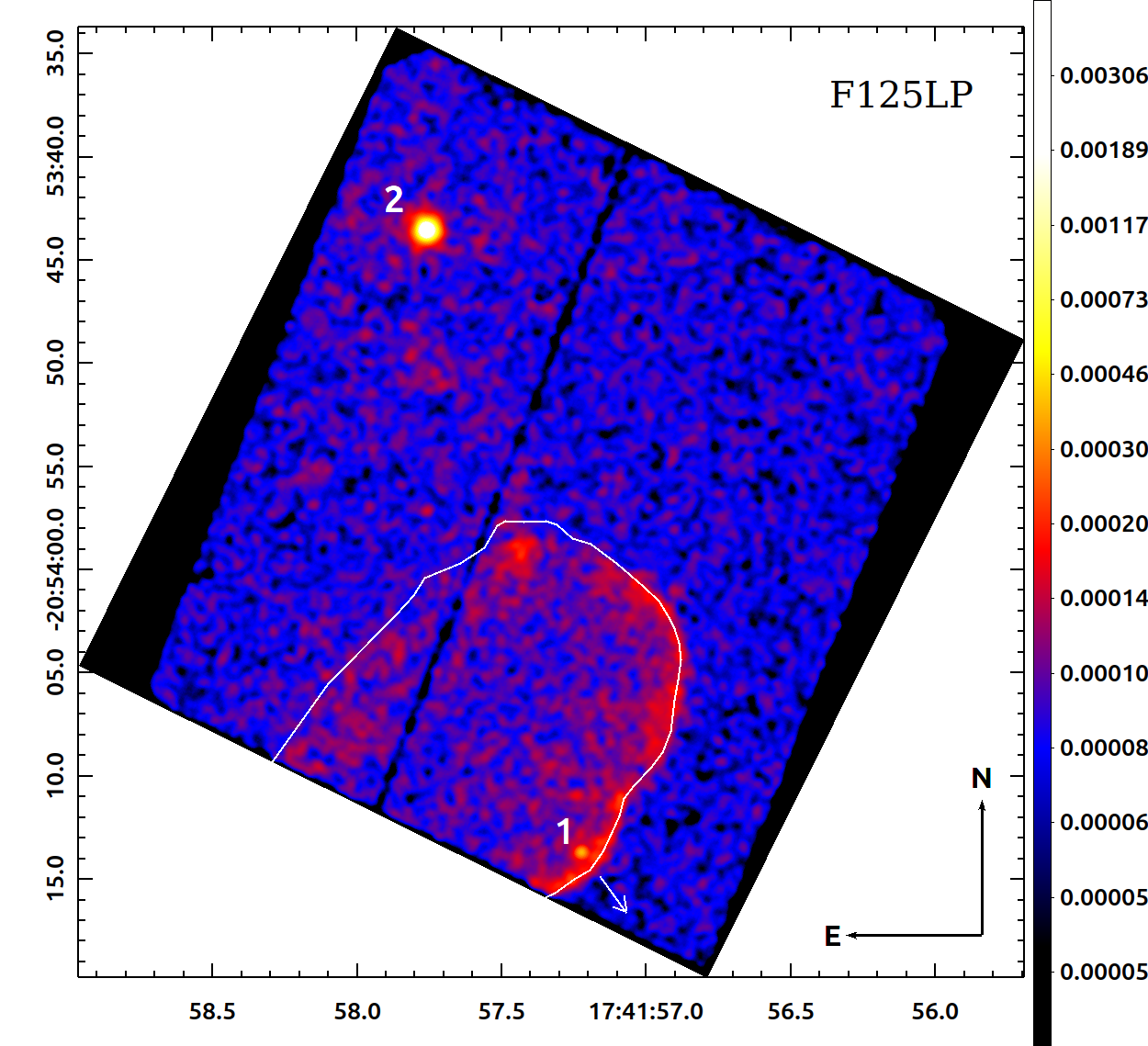}}
\end{minipage}
\hfill
\begin{minipage}[h]{0.5\linewidth}
\center{\includegraphics[width=0.85\linewidth]{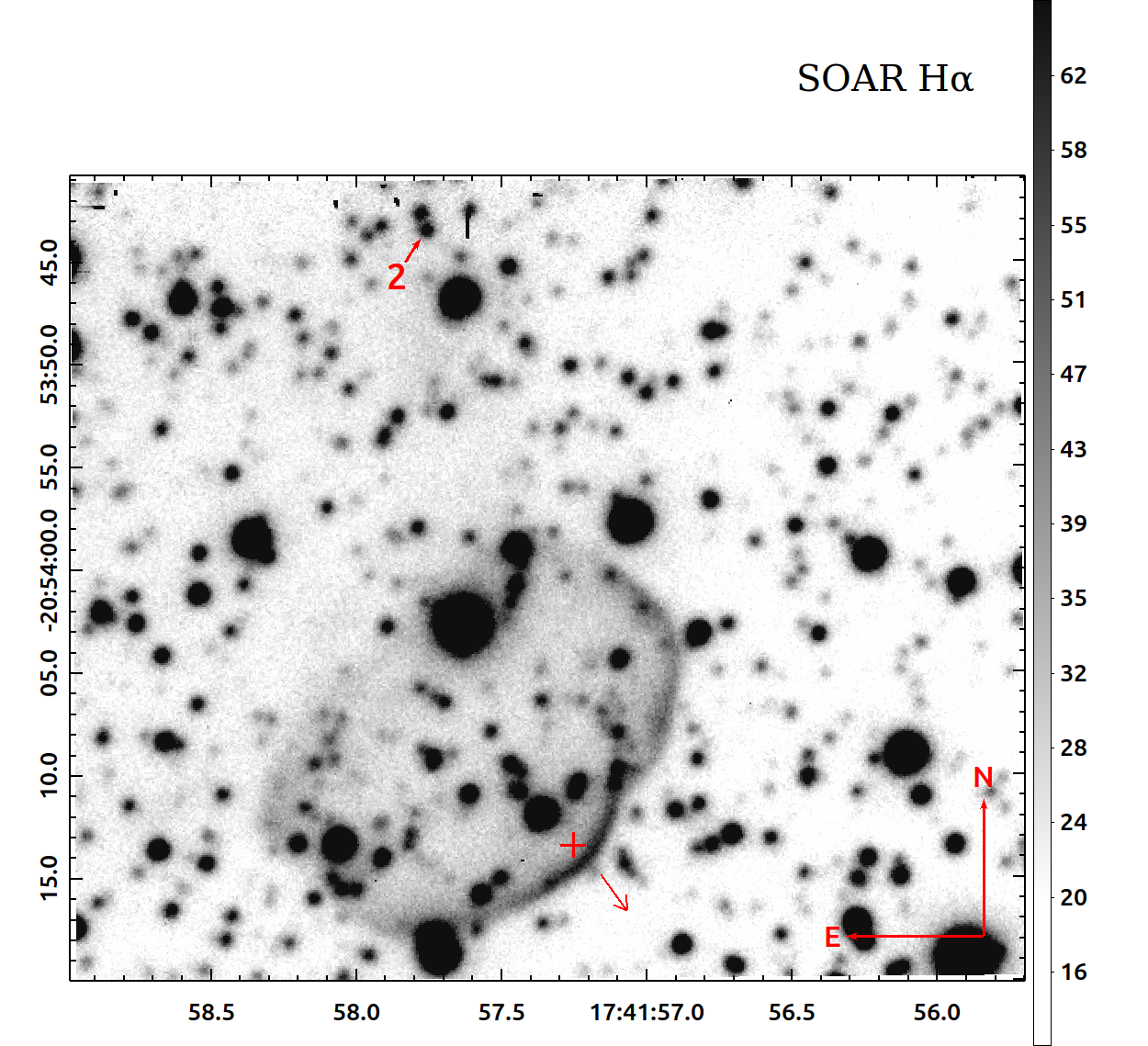}}
\end{minipage}
\hfill
\begin{minipage}[h]{0.5\linewidth}
\center{\includegraphics[width=0.85\linewidth]{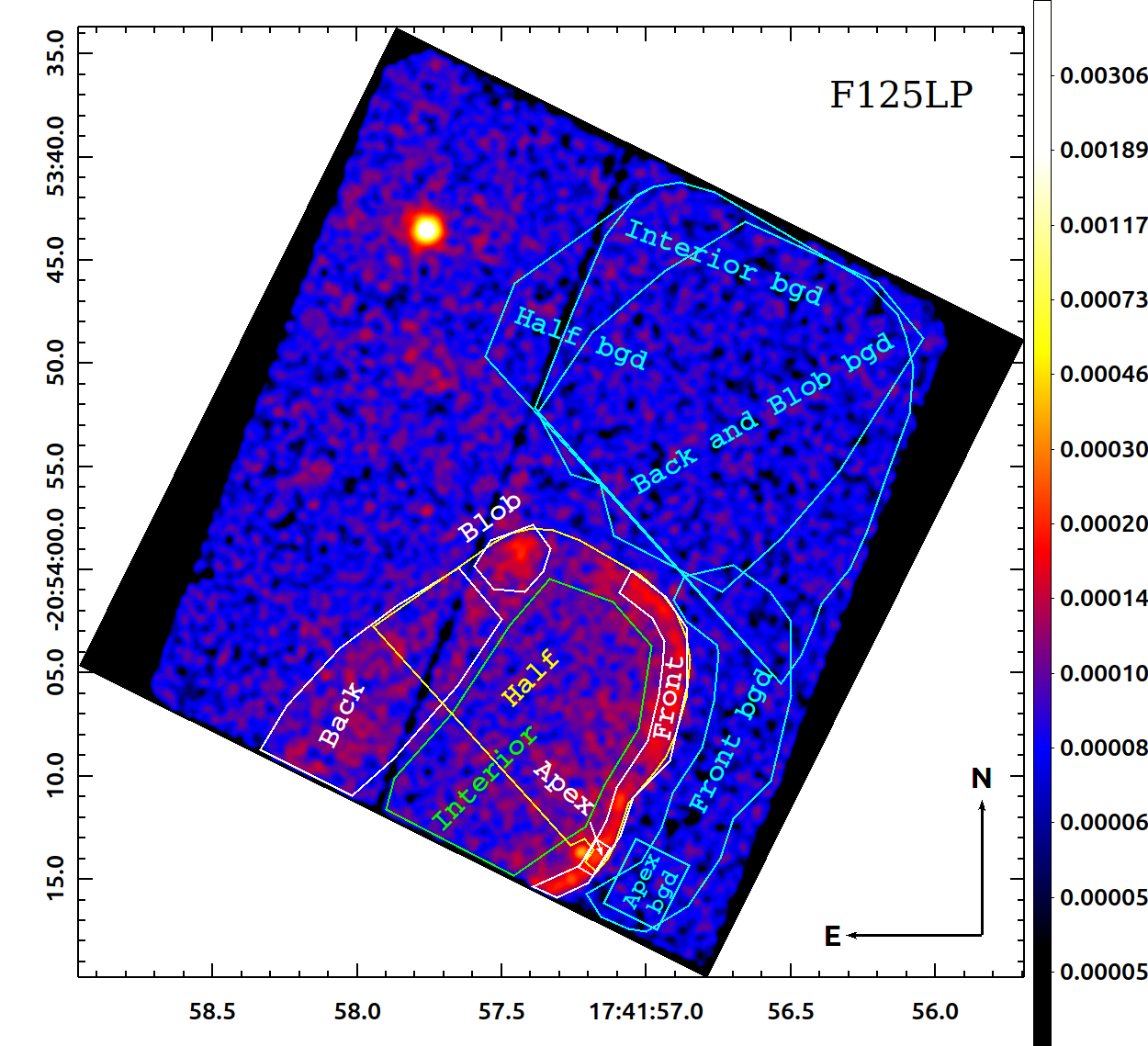}}
\end{minipage}
\hfill
\begin{minipage}[h]{0.5\linewidth}
\center{\includegraphics[width=0.85\linewidth]{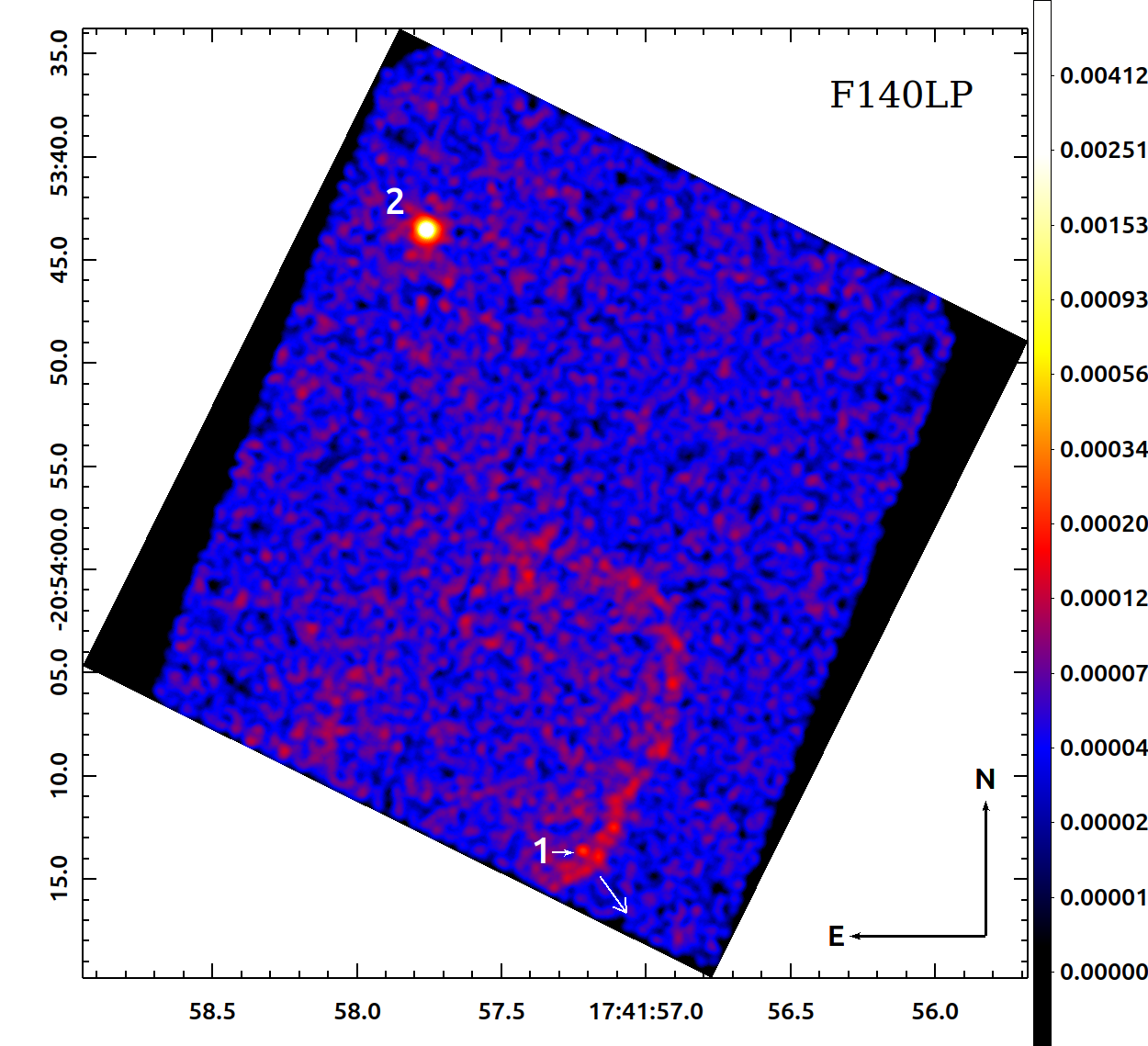}}
\end{minipage}
\caption{FUV and H$\alpha$ images of the field including J1741 and its BSN.
Source 1 is the pulsar counterpart, Source 2 is the bright UV point source, used for astrometry. 
The arrows show the direction of the pulsar's proper motion. 
Top left: F125LP image of the whole 
ACS/SBC FoV. The white 
contour shows the outer boundary of the H$\alpha$ 
BSN from the SOAR image. 
Top right: SOAR/SAM H$\alpha$ image of the 
BSN.  The red cross shows the pulsar position.
Bottom left: 
The same field as shown in the top-left panel. 
Six areas within the white, yellow and green contours were used for the flux measurements from the specific regions of the BSN (see Table~\ref{table:bsn_photometry}), 
the background for this measurement was extracted from the areas within the corresponding cyan contours. 
Bottom right: F140LP image. 
\label{fig:SBC_FoV}}
\end{figure*}


\section{Data analysis}

The SBC images in Figure~\ref{fig:SBC_FoV} show a prominent extended structure and two point sources marked by 1 and 2. The extended structure is obviously an FUV counterpart of the H$\alpha$ 
BSN. 
While 
the SBC images, centered on an outdated catalog position of J1741, 
do not fully cover the BSN,
the offset had the happy accident of including a UV-bright field star (Source 2), which has allowed us to refine the HST astrometry.
The archival H$\alpha$ BSN image, obtained on 2018 August 6 at the 4.1~m Southern Astrophysical Research (SOAR) telescope with SOAR Adaptive Optics Module (SAM)\footnote{
To the best of our knowledge, the image and its analysis have not been published. We took the data
from the NOIRLab Astro Data Archive 
and reduced them.} (PI Roger Romani), is shown in the top-right panel of 
Figure~\ref{fig:SBC_FoV}
(see also \citealt{Romani2010, Auchettl2015}). 

The fainter point source in Figure~\ref{fig:SBC_FoV} (Source 1 hereafter), located about 1\farcs0 -- 1\farcs2 
behind the bow shock apex, is the obvious candidate for the pulsar counterpart. 
As we show in the next subsection, this identification is supported by astrometric referencing of the HST FUV images.

\subsection{Astrometry}

The HST astrometric referencing
can be obtained   
with the aid of the brighter Source 2, seen in both the F125LP and F140LP images.  
Its positions in the FUV images are 
measured with a high precision
of about 1 mas. 

Source 2 was also detected in the near-UV by the {\em XMM-Newton} Optical Monitor (OM) \citep{Page2023} 
and the {\em Swift} Ultraviolet/Optical Telescope (UVOT) \citep{Yershov2015}. 
It likely corresponds to  {\em Gaia} DR3 source 4118158374505339904, with the ICRS coordinates $\alpha=17$:41:57.76052(4), $\delta=-20$:53:43.5218(4)\footnote{Numbers in parentheses denote 1$\sigma$ uncertainties relating to the last significant digits.}
(at  
MJD\,57388),
proper motion
$\mu_{\alpha} = -4.7 \pm 0.7$ mas yr$^{-1}$, 
$\mu_{\delta} = -0.7 \pm 0.4$ mas yr$^{-1}$, 
and magnitude $m_{G} = 19.88$. The SOAR H$\alpha$ image shows a very nearby ($\sim 0.8$\,arcsec) northern neighbour ({\em Gaia} DR3 source 4118158374505339776) that could, 
in principle, 
also be considered as the Source 2
counterpart. 
That 
northern neighbor, however, is much redder (detected in the near-IR but not detected in the near-UV with the {\em XMM-Newton} OM  and the {\em Swift} UVOT) and classified as a galaxy by \citet{Minniti2023}. 
Therefore, {\em Gaia} DR3 source 4118158374505339904 is 
the only probable counterpart of Source 2. 

With account for the proper motion,
the {\em Gaia} coordinates of the Source 2 differ from the  
coordinates in the F125LP image 
by 1\farcs09 
and 0\farcs19 
in R.A.\ and decl., respectively. This means that the WCS of the F125LP image  is offset from the {\em Gaia} WCS by about 
$1\farcs10$, which considerably exceeds the claimed HST pointing uncertainty of about 0\farcs1\footnote{See \url{https://hst-docs.stsci.edu/drizzpac/chapter-4-astrometric-information-in-the-header/4-5-absolute-astrometry}.}.
For the F140LP image, these offsets are 1\farcs32 in R.A.
and 0\farcs19 in decl.

We applied the corresponding boresight corrections to the F125LP and F140LP images 
and obtained the following Source 1 coordinates:
$\alpha=17$:41:57.221(1), $\delta=-20$:54:13.66(2) in the F125LP image 
and $\alpha=17$:41:57.218(1), $\delta=-20$:54:13.62(2) in the F140LP image.
However, being based on ``one-star astrometry'', these coordinates cannot be considered as precisely measured.

To prove that Source 1 is the pulsar counterpart, we compared these coordinates  with the coordinates  
of the optical  pulsar counterpart candidate 
detected by \citet{Mignani2016} with the  VLT FORS2 in the $b_{\rm high}$ and $v_{\rm high}$ bands. We re-reduced these VLT images,  
obtained in 2015.  
To get an accurate astrometric solution, we used 15 stars from the {\em Gaia} DR3 catalog \citep{GaiaCollaboration2016,Gaia2023}. The coordinates of the reference stars at the epoch of the VLT observations were calculated using the cataloged proper motion values and the default cataloged coordinates at the 2016.0 reference epoch. 
For the $b_{\rm high}$ image, we obtained the astrometric solution with uncertainties of 21 mas in R.A.\ and 20 mas in decl., using the IRAF tasks {\tt imcentroid} and {\tt ccmap}. This allowed us to measure the coordinates of the pulsar counterpart candidate in $b_{\rm high}$ image more accurately than it was done
by \citet{Mignani2016}.
Accounting for the centroiding uncertainties of the candidate, 
we obtained its 
coordinates 
 $\alpha =17$:41:57.268(3), $\delta =-20$:54:12.91(4) at the epoch of the VLT observations (MJD 57156). 
Applying the pulsar's proper motion \citep{Auchettl2015},
we obtain the predicted coordinates of the VLT candidate at the HST epoch (MJD 60107), $\alpha =17$:41:57.231(8) and $\delta =-20$:54:13.63(8), 
which are consistent with the position of the Source 1 within $\approx 1\sigma$ uncertainties.  
It means that the VLT counterpart candidate is the same object as the FUV candidate if (and only if) its  proper motion is the same as that of the pulsar. 
This virtually proves that the object is 
the pulsar counterpart.

\begin{figure*}[t]
\begin{minipage}[h]{0.5\linewidth}
\center{\includegraphics[width=0.75\linewidth]{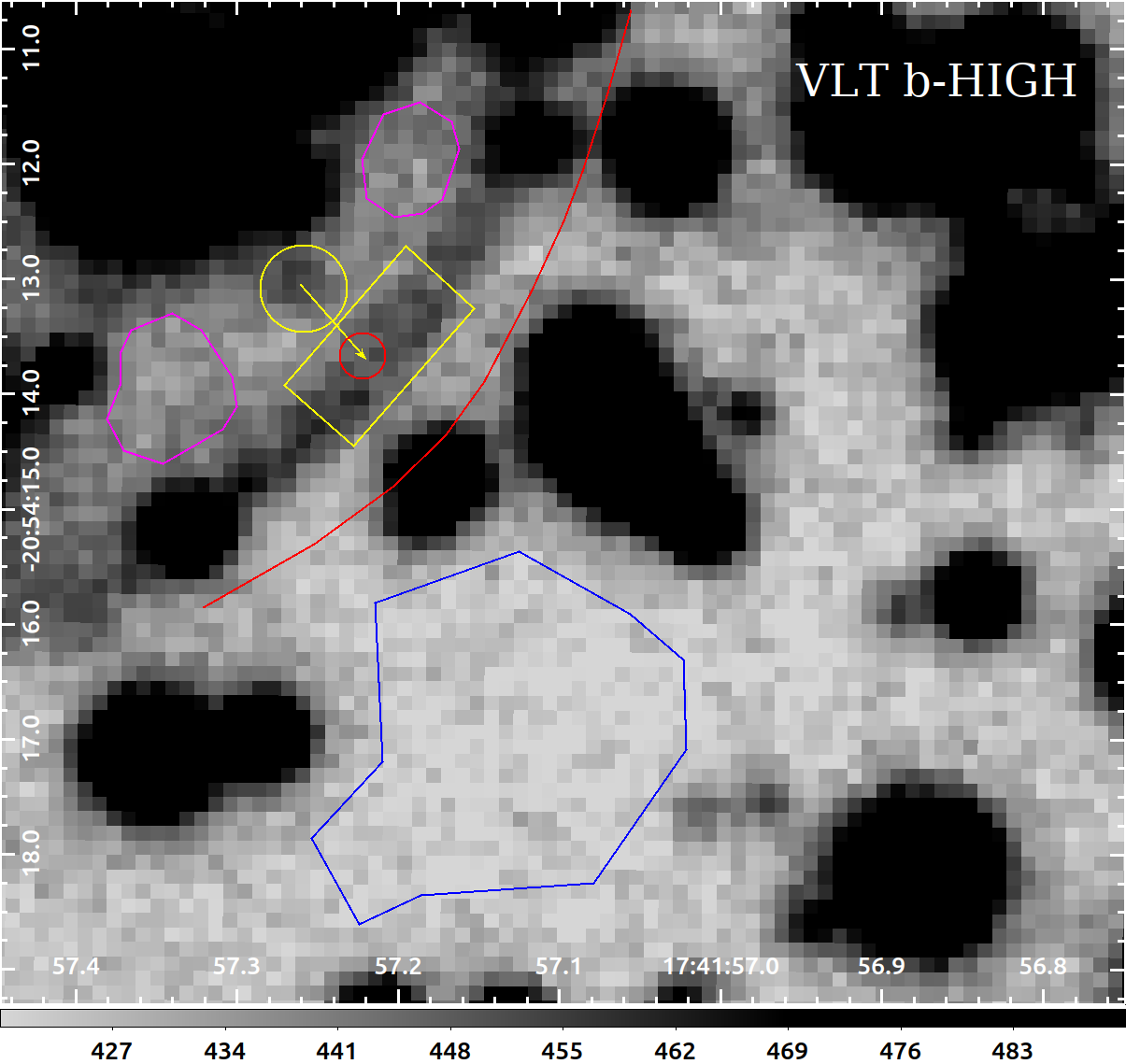}}
\end{minipage}
\hfill
\begin{minipage}[h]{0.5\linewidth}
\center{\includegraphics[width=0.75\linewidth]{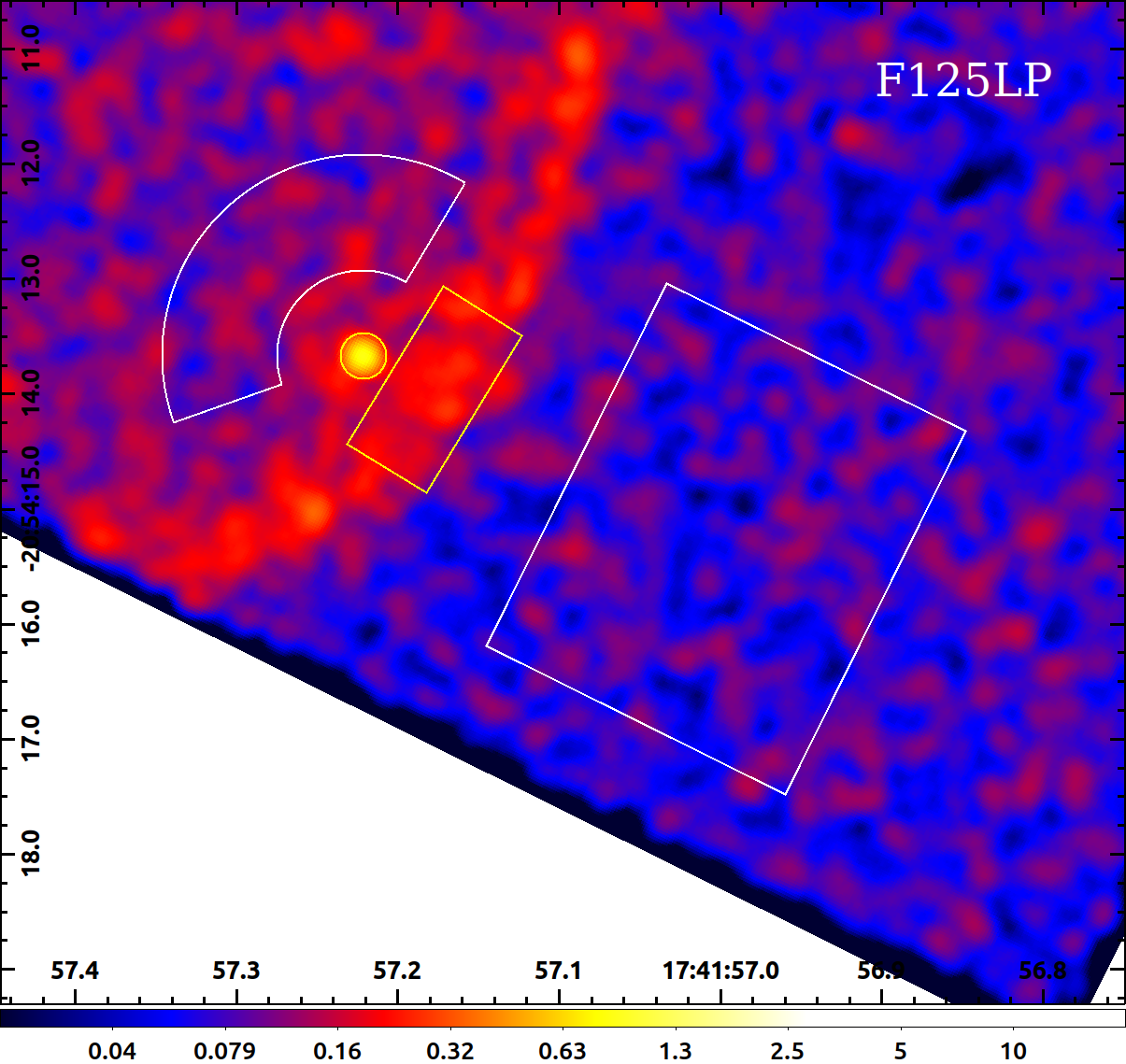}}
\end{minipage}
\caption{VLT $b_{\rm high}$ (MJD 57156) and HST F125LP (MJD 60107) images of the pulsar vicinity. 
The yellow rectangular regions were used for measurements of surface brightness of the 
BSN apex in the HST and VLT images. 
The 
yellow circles show apertures used for the pulsar counterpart flux measurements in the HST 
and VLT images, 
while the yellow arrow shows the 
proper motion shift of the pulsar between the epochs of 2015 and 2023. 
The red circle in the VLT image shows the pulsar counterpart position from the HST F125LP image.
The purple contours in the VLT image show the areas in which the background for pulsar counterpart was estimated, while the blue contour shows the background for the BSN apex.
The white segment of annulus and the white box in the F125LP image show the background area for the pulsar counterpart and the BSN apex,
respectively.
In all 
cases a set of 
apertures (circular or rectangular) was randomly generated inside the background area to estimate the mean and standard deviation of the background count rate. 
The red line in the VLT image shows the position of BSN outer boundary from the HST image.
\label{fig:F125LP_b-high}}
\end{figure*}

The proper-motion shift of the pulsar between the epochs of 2015 and 2023 is demonstrated in Figure~\ref{fig:F125LP_b-high}, which shows the pulsar vicinity in the F125LP and $b_{\rm high}$ images. 
Circles in the images show the positions of the counterpart 
candidate 
at the epochs of 2015 (yellow in the  left panel) and 2023 (red and yellow in the left and right panels, respectively).  
The  shift between these positions 
is 
1\farcs00 $\pm$ 0\farcs06, in agreement with the pulsar's 
proper motion shift  of 0\farcs88 $\pm$0\farcs12.

An additional argument for the detected optical-FUV source being the true pulsar counterpart is its location with respect to the 
BSN. 
The distance between it and the 
bow shock apex, moving to south-east,  
is about 1\farcs1 
in both the F125LP and $b_{\rm high}$ images. 
Thus, there remains no 
doubt that the optical-FUV source is indeed the pulsar counterpart.

\subsection{Photometry}
\subsubsection{Pulsar}
The pulsar counterpart count rates were measured in apertures of 8 pix (0\farcs2) and 10 pix (0\farcs25) radii in the F125LP and F140LP images 
(Table \ref{table:psr_photometry}),
which contain 62\% and 67\% of point source counts, respectively
\citep{Avila2016}. The background was extracted from the $135^\circ$ sector of the annulus with inner 
and outer radii of 30 pix (0\farcs75) and 
70 pix (1\farcs75) located behind the bow shock (right panel of Figure \ref{fig:F125LP_b-high}) in both SBC images. 
Because of 
proximity of 
bright background stars in 
the optical images, small apertures of 3 pix (0\farcs38) 
and 2 pix (0\farcs25) 
were used in the 
$b_{\rm high}$ and $v_{\rm high}$ images, respectively. The aperture correction in this case was determined with the aid of bright unsaturated stars. The background 
was extracted from the two small polygons behind the bow shock in the VLT images (left panel of Figure~\ref{fig:F125LP_b-high}). 
We have to note that limited choice of 
background regions in the very crowded field could lead to an additional systematic error.

To evaluate the count rate uncertainties,
we used the "empty aperture" approach \citep{Skelton2014,Abramkin2022}.  
The background count 
rates were measured in multiple apertures of the same shape and size as the source apertures, uniformly distributed in the background regions. We estimated the mean $\overline{C}_{\rm bgd}$ and the variance $\sigma_{C_{\rm bgd}}^2$ of the count rates inside the aperture, and calculated the net source count rate $C_{s}$ and its 
uncertainty $\sigma_{s}$:
\begin{equation}
    C_s = C_{\rm tot} -\overline{C}_{\rm bgd},\quad\quad \sigma_s = \left(\sigma_{C_{\rm bgd}}^2 + C_st_{\rm exp}^{-1} g^{-1}  
    \right)^{1/2},
\end{equation}
where $C_{\rm tot}$ is the total count rate in the source aperture, $t_{\rm exp}$ is the exposure time, and $g$ is the gain ($g=1$ and 1.25 for the FUV and optical data, respectively).
In the SBC images, we used a set of 1000 background apertures. 
For VLT photometry we used 50 and 35 background apertures in $b_{\rm high}$ and 
$v_{\rm high}$ images, respectively.

Photometric calibration for the optical $b_{\rm high}$ and $v_{\rm high}$ filters 
was applied using the mean night instrumental zeropoints 
and the atmospheric extinction 
coefficients\footnote{See \url{https://archive.eso.org/qc1/qc1_cgi?action=qc1_browse_table&table=fors2_photometry}.}. 
Our $v_{\rm high}$ and $b_{\rm high}$ 
pulsar fluxes are slightly lower than
those obtained by \citet{Mignani2016}, 
while our uncertainties are significantly larger (see Table~\ref{table:psr_photometry}).

To convert the 
aperture-corrected pulsar's count rate measured in a given SBC filter $i$, $C_{s,i}/\phi_i$ (where $\phi_i$ is the fraction of point source counts in the chosen aperture) to the mean flux density in that filter, 
\begin{equation}
   \langle f_\nu\rangle_i \equiv 
   \frac{\int f_\nu(\nu) T_i(\nu) \nu^{-1}\,d\nu}{\int T_i(\nu) \nu^{-1}\,d\nu} =
   \frac{\int f_\nu(\lambda) T_i(\lambda)\lambda^{-1}\, d\lambda}
    {\int T_i(\lambda) \lambda^{-1}\, d\lambda}
    = \frac{C_{s,i}}{\phi_i}{\cal P}_{\nu,i}\,,
\label{eq:mean_flux}
\end{equation}
we used the inverse sensitivity ${\cal P_\nu}$ 
obtained from the header keywords {\tt photflam} and {\tt photplam} in the data files. Note that the mean flux defined by Equation (\ref{eq:mean_flux}) is fully determined by the measured count rate and filter transmission $T_i(\lambda)$ (see Figure~\ref{fig:filters}),
and it is not tied to a specific wavelength within the filter passband. It can be 
related to the ``isophotal wavelength'' $\lambda_{\rm iso}$ \citep{Tokunaga2005}, defined by the equation $f_\nu(\lambda_{\rm iso}) = \langle f_\nu \rangle$, but $\lambda_{\rm iso}$ does depend on the spectral shape, and it cannot be determined from a count rate measurement in just one filter (unless the filter passband is very narrow). 
Moreover, the unique value of $\lambda_{\rm iso}$ can only be found if $f_\nu(\lambda)$ varies monotonously within the passband (as we expect for a pulsar FUV spectrum but not necessarily for a BSN spectrum).



\begin{table*}
\setlength{\tabcolsep}{0.75em}
\caption{Pulsar 
photometry}             
\label{table:psr_photometry}      
\centering          
\begin{tabular}{l c c c c c c D{,}{\,\pm\,}{-1} D{,}{\,\pm\,}{-1} c D{,}{\,\pm\,}{-1}}     
\hline\hline       
\rule{0pt}{2ex} Filter & $\lambda_{\rm p}$ & $W_{\rm eff}$ & $t_{\rm exp}$ & $r_{\rm extr}$ & $\phi$ & $C_{\rm tot}$ & 
\multicolumn{1}{c}{$C_{\rm bgd}$}
& \multicolumn{1}{c}{$C_{s}$} & ${\cal P}_\nu$ & \multicolumn{1}{c}{$\langle f_{\nu}\rangle$} \\
\rule{0pt}{2.5ex} & \AA & \AA & s & arcsec &  \%
& cnts/ks & \multicolumn{1}{c}{cnts/ks} & \multicolumn{1}{c}{cnts/ks} & 
nJy\,ks/cnts & \multicolumn{1}{c}{nJy} \\

\hline              
\rule{0pt}{2.2ex}F125LP & 1426 & 340 & 1830 & 0.2 & 62 & 103.8 & 23.1,3.7  & 80.7,7.6 & 0.875 & 113,11\\
F140LP & 1519 & 259 & 566 & 0.25 & 67 &  77.0 & 17.7,6.6  & 59.3,12.2 & 1.668 & 148,30\\
$b_{\rm high}$ & 4360 & 952 & 4680 & 0.38 & 70 & 71200 & 69420,370 & 1780,370 & \ldots  & 85,18\\
$v_{\rm high}$ & 5524 & 1187 & 5400 & 0.25 & 51 & 97970 & 94560,920  & 3410,920 & \ldots & 163,44 \\[0.25ex]
\hline                
\end{tabular}
\tablefoot{
$\lambda_{\rm p}$ and $W_{\rm eff}= \left[\int T(\lambda) d\lambda\right]\left[{\rm Max}(T(\lambda))\right]^{-1}$ are 
the pivot wavelength 
and the
effective width of the filter, 
$\phi$ is the fraction of source counts in the aperture with radius $r_{\rm extr}$,
$C_{\rm tot}$ is the total count rate in the source aperture,
$C_{\rm bgd}=\overline{C}_{\rm bgd}\pm\sigma_{C_{\rm bgd}}$, 
$\overline{C}_{\rm bgd}$ and $\sigma_{C_{\rm bgd}}$ are the mean and standard deviation of background measurements,
$C_{s}$ is the net source count rate, its error is estimated as 
$[(\sigma_{C{\rm bgd}}^2 + C_{s} t_{\rm exp}^{-1} g^{-1}]^{1/2}$, where $g$ is the gain, 
$\langle f_\nu \rangle = C_{s} {\cal P}_\nu$/$\phi$
is the mean flux density, 
and ${\cal P}_\nu$ is the count rate-to-flux conversion factor.
}
\end{table*}

\subsubsection{Bow shock nebula}

To explore the  
{\em BSN} emission in the 
F125LP, F140LP and $b_{\rm high}$ images, we 
measured the count rates in six BSN regions, which we call Apex, Front, Back, Blob, Interior and Half. These BSN regions and the corresponding background regions are shown in Figures \ref{fig:SBC_FoV} and \ref{fig:F125LP_b-high}, while their areas are presented in Table~\ref{table:bsn_photometry}.

The brightest Apex region is a tiny  $0\farcs8\times1\farcs6$ 
rectangle
near the BSN apex  
(shown by the yellow rectangular contours in Figure \ref{fig:F125LP_b-high}).
 The background for 
 this measurement was extracted from a $3\farcs6\times2\farcs8$
 rectangle ahead of the bow shock in the 
FUV images, where 1000 rectangular background apertures were generated applying the same "empty aperture" approach that we used for the pulsar photometry. In the very crowded 
$b_{\rm high}$ field we extracted the background  
from a 6.9 arcsec$^2$ polygon 
(see the left panel of Figure \ref{fig:F125LP_b-high}), where 85 background apertures were generated. Because of the crowded 
field, we were not able to accurately measure the count rates from the other BSN regions of the $b_{\rm high}$ field. 

The Front region 
includes the part of the flattish front of the BSN shell that was imaged onto the SBC detector.
Some diffuse emission is seen throughout the entire ellipse-like BSN, with brightenings in its back part and in a compact region in the northern extreme of the BSN -- the Back and Blob regions.
The Interior region 
inside the closed BSN shell image possibly includes a contribution from the inner part of the 3D BSN. 
Finally, we measured the 
count rate from a half of the whole ellipse-like nebula seen in the FUV images (the Half region).

For the BSN photometry in the SBC filters, we measured the source count rates $C_s$ for the six BSN regions as described above and, in addition, calculated mean surface brightnesses ${\cal B}_s = C_s/A_s$ (see Table~\ref{table:bsn_photometry}).
For the 
BSN FUV photometry in the Front, Interior and 
Half regions, we used sets of 20, 100 and 150 
background apertures, 
respectively, placed in the corresponding background areas (see 
the bottom left panel of Figure~\ref{fig:SBC_FoV}). For the Back and Blob regions, we 
used sets of 50 background apertures placed in the same 
background area. 
We converted the source count rates to the mean flux densities $\langle f_\nu\rangle$ using Equation (\ref{eq:mean_flux}) with $\phi=1$. If the BSN spectrum is a monotonous continuum, then there exists one isophotal wavelength within the filter passband, as described above. If, however, the BSN spectrum is dominated by spectral 
lines, then $f_\nu(\lambda) =\langle f_\nu\rangle$ at several values of $\lambda$, and the concept of the mean flux density becomes less useful. 

\begin{table*}
\setlength{\tabcolsep}{0.5em}
\caption{BSN photometry}             
\label{table:bsn_photometry}      
\centering          
\begin{tabular}{lcccc D{,}{\,\pm\,}{-1} D{,}{\,\pm\,}{-1} D{,}{\,\pm\,}{-1} D{,}{\,\pm\,}{-1} D{,}{\,\pm\,}{-1}}     
\hline\hline                         
\rule{0pt}{2ex} Region & Filter &  $A_{\rm s}$ & $A_{\rm bdg}$ &  $C_{\rm tot}$ & \multicolumn{1}{c}{$C_{\rm bgd}$}
& \multicolumn{1}{c}{$C_{s}$} & \multicolumn{1}{c}{${\cal B}_s$} & \multicolumn{1}{c}{$\langle f_{\nu} \rangle$} & \multicolumn{1}{c}{$\langle I_\nu\rangle$} \\
\rule{0pt}{2.5ex} & & arcsec$^{2}$  & arcsec$^2$ 
& cnts/ks & \multicolumn{1}{c}{cnts/ks} & \multicolumn{1}{c}{cnts/ks} & \multicolumn{1}{c}{$\frac{\rm cnt}{{\rm ks\, arcsec}^2}$} & \multicolumn{1}{c}{nJy} & \multicolumn{1}{c}{$\frac{\rm nJy}{\rm arcsec^2}$} \\[1ex]
\hline   
\rule{0pt}{2.2ex} Apex & F125LP &  1.28 & 10.1 & 392 & 155,14 & 237,18 & 185,14 & 207,16 & 162,12\\
                       & F140LP &  1.28 & 10.1 &264 & 85,18 &  179,25 & 140,20 & 298,42 & 233,33\\
 & $b_{\rm high}$ &  1.28 & 6.9 & 201900 & 190000,1200 &  11900,1200 & 9300,900 & 400,40 & 310,30\\
\rule{0pt}{3ex} Front & F125LP &  20.2 & 68.1 & 4719 & 2502,28 & 2217,44 & 110,2 & 1940,40 & 96,2\\
                      & F140LP &  20.2 & 68.1  &2823 & 1422,63 & 1401,80 & 69,4 & 2340,130 & 116,7\\
\rule{0pt}{3ex} Back & F125LP &  59.8 & 195  &9406 & 7708,34 & 1698,46 & 28.4,0.8 & 1490,40 & 25,1\\
                     & F140LP &  59.8 & 195 & 5568 & 4467,59 & 1101,74 & 18.4,1.2 & 1840,120 & 31,2\\
\rule{0pt}{3ex} Blob & F125LP & 7.9 & 195 & 1597 & 1028,27 & 569,32 & 72,4 & 500,30 & 63,4\\
                     & F140LP & 7.9 & 195 & 1018 &  590,38 & 428,47 & 54,6 & 710,80 & 90,10\\
\rule{0pt}{3ex} Half & F125LP & 142.7 & 292 & 24722 & 18356,103 & 6366,119 & 44.6,0.8 & 5570,100 & 39,1\\
                     & F140LP & 142.7 & 292 & 14530 & 10590,108 & 3940,137 & 27.6,1.0 &  6570,230 & 46,2\\
\rule{0pt}{3ex} Interior & F125LP & 105.1 & 267 & 16596 & 13546,78 & 3050,90 & 29.0,0.9 & 2670,80 & 25,1\\
                         & F140LP & 105.1 & 267 & 9340  & 7823,91 & 1517,105 & 14.4,1.0 & 2530,180 & 24,2\\
\hline
\end{tabular}
\tablefoot{$A_{\rm s}$ and $A_{\rm bgd}$ are the source extraction area 
and the background area for the BSN regions.
${\cal B}_s = C_s/A_{\rm s}$ is the average surface brightness of the BSN in a given region. 
$\langle I_\nu\rangle = \langle f_\nu\rangle/A_s$ is the average specific intensity.}
\end{table*}

\section{Model fits}
\subsection{Pulsar Optical-FUV Spectral Fits}
We 
first fit the optical-FUV data with the simple absorbed PL model
\begin{equation}
    f_\nu^{\rm mod} = f_{\nu,0}\,
    (\nu/\nu_0)^\alpha\times 10^{-0.4 A(\nu)}\,,
\label{eq:PL}
\end{equation}
where $\nu_0$ and $\lambda_0$ are the reference frequency and wavelength (we chose $\nu_0 = 10^{15}$ Hz, which corresponds to $\lambda_0 = 3000$\,\AA). 
The interstellar extinction 
$A(\nu)$ (or $A(\lambda)$) is proportional to the color excess $E(B-V)$. 
For the DM-distance of 270 pc, resulting from the Galactic electron distribution 
model by \cite{Yao2017}, the  recent 
 extinction map by  \cite{Edenhofer2024} gives $E(B-V)=0.22$. We also checked \cite{Lallement2022}, \cite{Vergely2022} and Bayestar19 \citep{Green2019} extinction maps and obtained the similar values. We will 
 consider $E(B - V) \simeq 0.22$ as the most 
probable 
color excess for $d = 270$ pc and use \citet{Cardelli1989} to connect 
$A(\lambda)$ with the color excess. 
For this $E(B - V)$,                       
the wavelength dependence of the extinction coefficient $10^{-0.4 A(\lambda)}$
in the FUV range is shown in Figure~\ref{fig:filters}.

We vary two model parameters (PL slope $\alpha$ and normalization $f_{\nu,0}$) at a fixed value of the  color excess to minimize the following $\chi^2$ statistic:
\begin{equation}
    \chi^2 = \sum_i \frac{\left(\langle f_{\nu}\rangle_{i} - \langle f_{\nu}^{\rm mod}\rangle _{i}\right)^2}{\sigma_{\langle f_{\nu}\rangle_i}^2}\,,
\end{equation}
where $\langle f_{\nu}\rangle_{i}$ and $\sigma_{\langle f_{\nu}\rangle_i}$ are the mean flux density and its error in the $i$-th filter; the mean model flux is given by Equation~(\ref{eq:mean_flux}) in which $f_\nu$ is replaced by $f_\nu^{\rm mod}$.

\begin{figure*}[t]
\begin{minipage}[h]{0.515\linewidth}
\center{\includegraphics[width=0.85\linewidth]{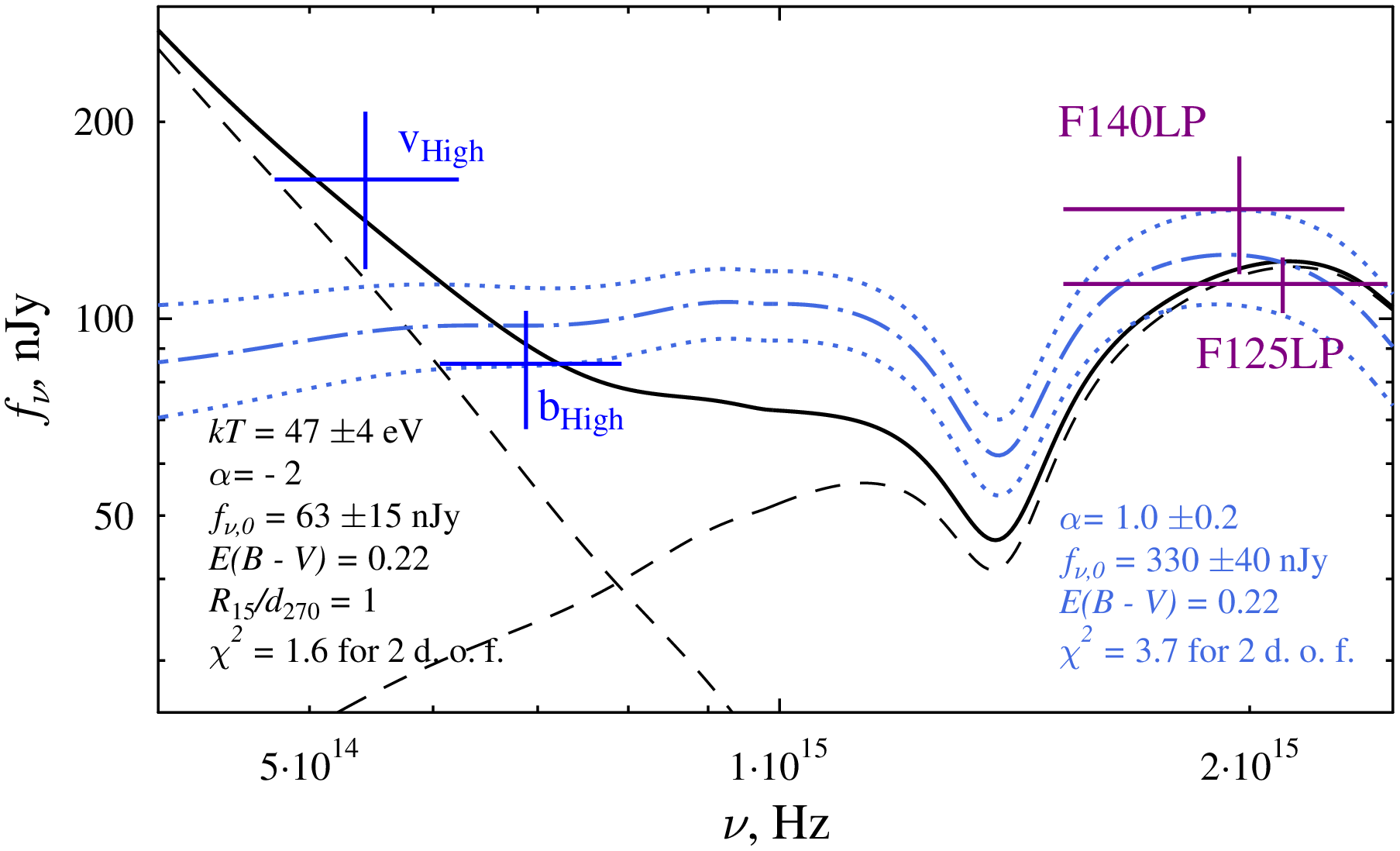}}
\end{minipage}
\hfill
\begin{minipage}[h]{0.485\linewidth}
\center{\includegraphics[width=0.85\linewidth]{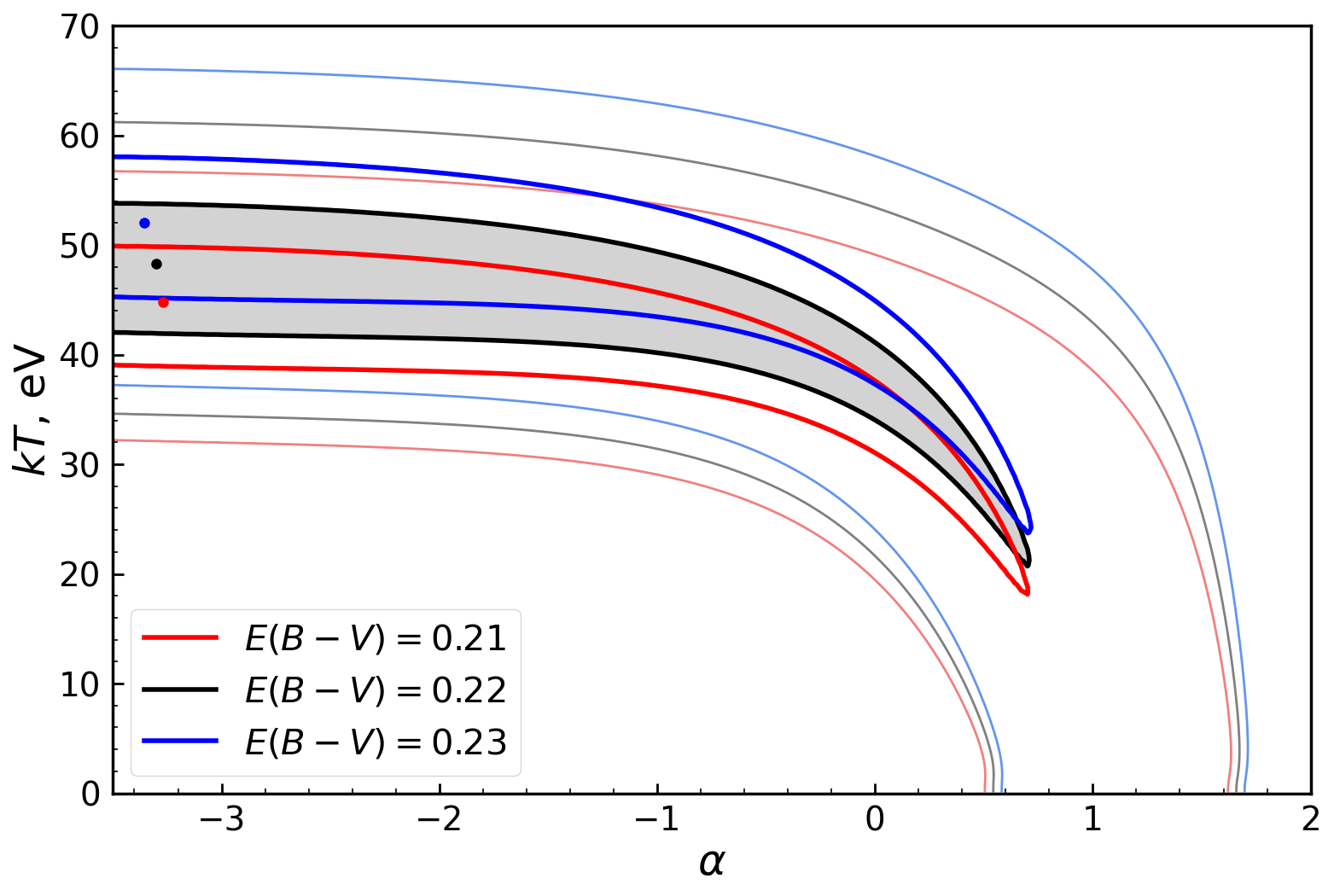}}
\end{minipage}
\caption{
Left: Fits of 
the optical-FUV spectrum. 
The 
dash-dotted and dotted blue lines show the best PL fit 
and its 68\% uncertainties. 
The solid and dashed black lines show an example of a PL+BB fit 
with fixed $\alpha = -2$, $R_{15}/d_{270} = 1$ at $E(B-V)= 0.22$ and its components. 
Right: 
Confidence contours in the $\alpha$--$kT$ plane for $R_{15}/d_{270} = 1$, at 68.3\%, 99.7\% confidence levels (for two parameters of interest). 
\label{fig:bestfit_conf}}
\end{figure*}

For the absorbed PL model, we obtained $\alpha = +1.0 \pm 0.2$ (i.e., photon index $\Gamma = 1-\alpha = 0.0 \pm 0.2$), 
$f_{\nu,0} = 330 \pm 40$ nJy, 
with $\chi^{2} = 3.7$ 
for 2 
degrees of freedom (dof), for the color excess $E(B - V) = 0.22$. 
The best-fit spectrum and its uncertainties are shown by dash-dot and dotted blue lines in the left panel of Figure~\ref{fig:bestfit_conf}. Not only the fit is 
formally unacceptable, but also such large positive PL indices 
have not been seen in any other pulsar in the optical-UV. 

Therefore, we can assume that the FUV flux is, at least partly, due to thermal emission.  
We thus fit the optical-FUV flux densities with the PL+BB model 
\begin{equation}
    f_\nu^{\rm mod} =\left[f_{\nu,0} \left(\frac{\nu}{\nu_0}\right)^\alpha + \frac{R^2}{d^2} \pi B_\nu(T)\right]\times 10^{-0.4 A(\nu)},
\label{eq:PL+BB}
\end{equation}
where $d=270 d_{270}$ pc is the distance,
$T$ and  $R$ are the NS temperature and radius as seen by a distant observer, and $B_\nu(T)$ is the Planck function.
The best-fit parameters for the BB+PL model are 
$kT = 48 \pm 4$ eV,
$\alpha = -3.3 \pm 2.2$ (i.e., $\Gamma=4.3 \pm 2.2$), $f_{\nu, 0} = 33^{+70}_{-20}$ 
nJy at the fixed values of $E(B - V) = 0.22$ and $R_{15}/d_{270} = 1$ (where 
$R = 15 R_{15}$ km) with $\chi^{2} = 1.2$ for 
1 dof.

The right panel of Figure \ref{fig:bestfit_conf} shows the 68.3\%, and 99.7\% confidence contours 
(for two parameters of
interest) for $R_{15}/d_{270} = 1$ and three values of the color excess: $E(B - V) = 0.21$, 0.22, and 0.23. We see that the fits are almost insensitive to the spectral slope value at $\alpha \lesssim -2$, 
because of few data points. Therefore, we tried to fit the BB+PL model with the 
spectral slope fixed at $\alpha = -2$. In this case, we obtained the following best-fit parameters: 
$kT=47 \pm 4$ eV,
$f_{\nu, 0} = 63 \pm 15$ nJy, $\chi^{2} = 1.6$ for 2 dof at $R_{15}/d_{270} = 1$, for $E(B - V) = 0.22$ 
(see the left panel of Figure~\ref{fig:bestfit_conf}). 

For an alternative DM-distance of 380 pc, resulting from the Galactic electron distribution model by \cite{Cordes2002}, the  extinction map by \cite{Edenhofer2024} suggests a larger color excess,  $E(B-V)=0.26$. 
The larger distance and extinction lead to
unrealistically  high  temperatures and/or thermally emitting areas  derived from the optical-FUV fits, 
inconsistent with those obtained in X-rays. For instance, we got $T\sim120$ eV for $R_{15}/d_{380} \sim 1$ 
versus $T\sim 60$ eV  and   $R_{15}/d_{380} \sim 0.3$ in X-rays (See Sect.~\ref{intro}). We thus adopt 
$d=270$ pc and $E(B-V)=0.22$ for 
further analysis.

\subsection{Optical through X-ray Spectrum of the Pulsar}

The pulsar surface temperature derived from the optical-FUV spectral fit 
is generally compatible with that obtained from the {\em Chandra} X-ray spectra using the same BB+PL emission model \citep[][]{Karpova2014,Auchettl2015}. However, continuation of the PL component of this model obtained 
from the X-ray data 
alone  overshoots the optical-FUV fluxes  by several orders of 
magnitude.
It means that the optical through X-ray nonthermal emission of the pulsar cannot be described by a single PL, but it can possibly be described by a broken PL (brPL) model:
\begin{equation}
    \frac{dN}{dE} = K_{\rm PL} (E_{\rm brk}/E_0)^{-\Gamma_{\rm opt}} \left\{
      \begin{array}{lr}
          (E/E_{\rm brk})^{-\Gamma_{\rm opt}} & \quad \text{if}\; E \leq E_{\rm brk}\\
         (E/E_{\rm brk})^{-\Gamma_{\rm X}} & \quad \text{if}\; E > E_{\rm brk}
      \end{array}\right.\,,
\end{equation}
%
where $E_{0}=1$ keV is the reference energy, $E_{\rm brk}$ is the break energy, $\Gamma_{\rm opt} = 1 - \alpha_{\rm opt}$ and 
$\Gamma_{\rm X} = 1 - \alpha_{\rm X}$ are photon indices at energies lower and higher than $E_{\rm brk}$, and the normalization $K_{\rm PL}$ is the photon flux density at $E=E_0$ for the low-energy branch (or its extrapolation).

To check this hypothesis and 
estimate the nonthermal and thermal model parameters,
we fitted, using XSPEC spectral fitting package (ver.\ 12.13.0, \citealt{Arnaud1996}), the FUV and optical data points together with the X-ray spectra obtained with the {\em Chandra} ACIS-S instrument (programs ``A Legacy Study of the Relativistic Shocks of PWNe'', PI Roger Romani, and ''Constraining the Distance \& Temperature of LAT PSR J1742--20, The Newly Discovered Nearby Middle-Aged Neutron Star'', PI  Gregory Sivakoff) and reduced by \cite{Karpova2014}. 
In our fits, we used the T\"{u}bingen-Boulder interstellar medium (ISM) absorption model ({\sc tbabs}) 
with Wilms abundances \citep{Wilms2000} in X-rays and the {\sc redden} interstellar extinction law \citep{Cardelli1989} in the optical-UV. 
Following \citet{Zharikov2021} and \citet{Zyuzin2022}, we used the {\tt python} package {\tt emcee} \citep{Foreman-Mackey2013} to fit the data with the Markov chain Monte Carlo (MCMC) approach, using the affine invariant MCMC sampler \citep{Goodman2010}. We utilized 1000 walkers and 50000 steps for each of the tried models. 

\begin{figure*}[t]
\begin{minipage}[h]{0.5\linewidth}
\center{\includegraphics[width=0.9\linewidth]{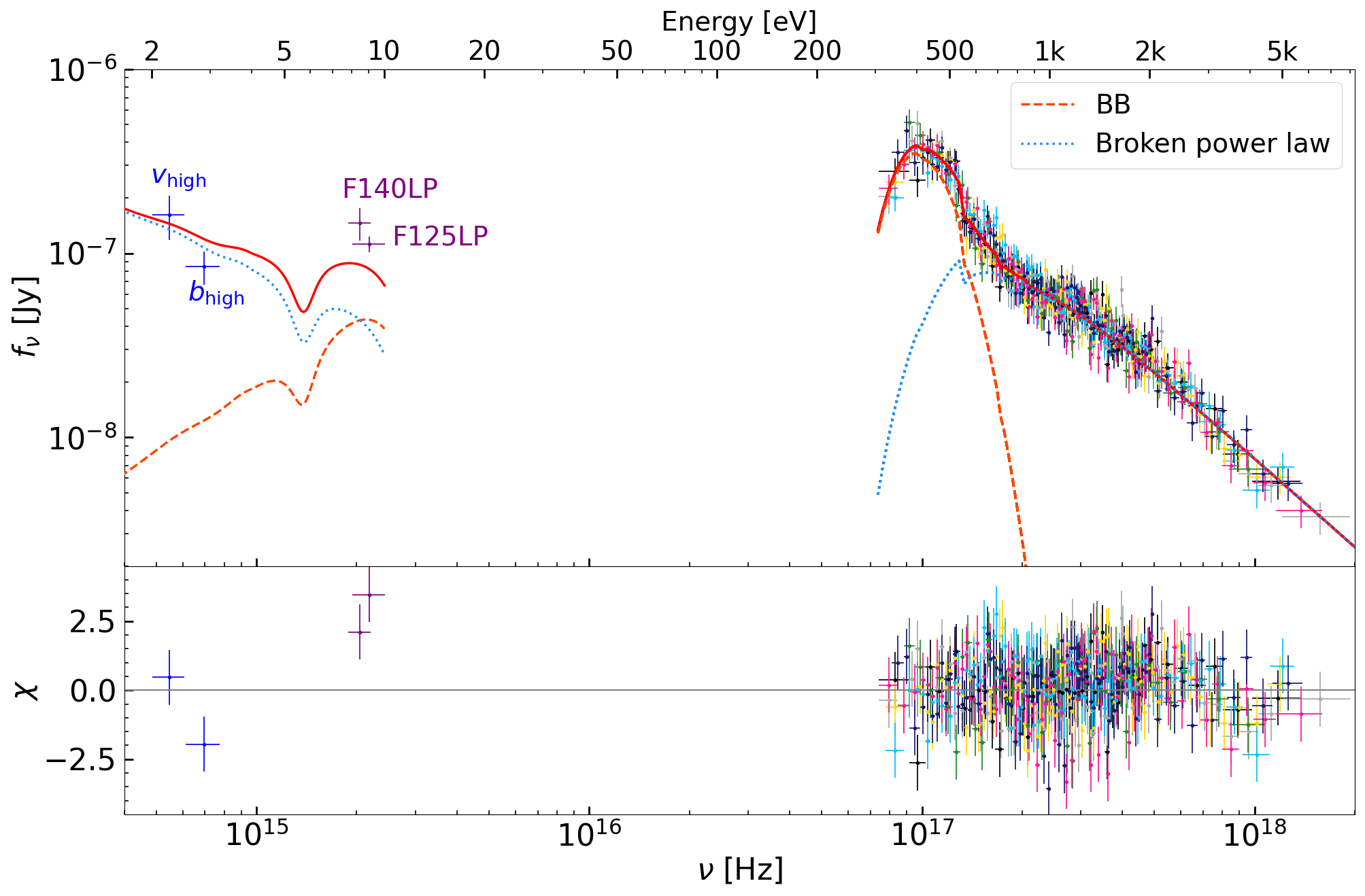}}
\end{minipage}
\hfill
\begin{minipage}[h]{0.5\linewidth}
\center{\includegraphics[width=0.9\linewidth]{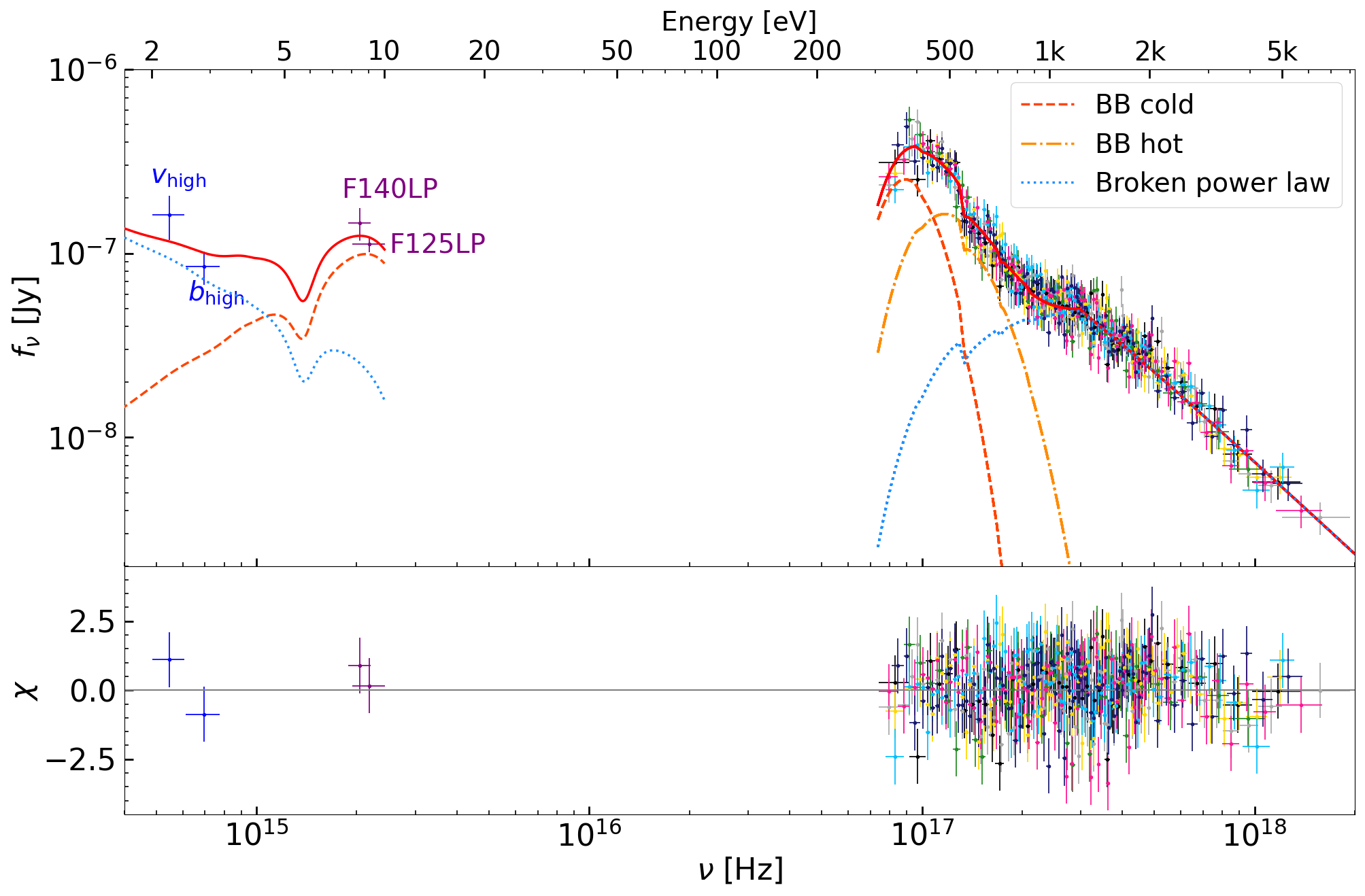}}
\end{minipage}
\caption{Observed 
optical-FUV and X-ray spectra of J1741 fitted 
with the absorbed 
brPL + BB (\textit{left}) and brPL + BB + BB (\textit{right}) models.
The blue and violet data points show the VLT optical and HST FUV data, respectively, while the seven separate X-ray data sets from 
{\em Chandra} observations (see Table 1 in \cite{Auchettl2015}) are shown by different colors.
The solid red lines show the best-fit models (see text and 
Figure~\ref{fig:triangle_optxrays_bbbb+bknpl} for the best-fit parameters), whereas the red dashed, orange dot-dashed and blue dotted lines show model components, as indicated in  legends.  
The fit residuals shown in the bottom sub-panels are defined as $\chi\equiv({\rm data} - {\rm model})/{\rm error}$.
\label{fig:opt-xrays_fit}}
\end{figure*}

We used the hydrogen column density value, $N_{\rm H,21}=1.20^{+0.08}_{-0.07}$, measured by \cite{Auchettl2015} 
from the PL spectral fit of the compact PWN in a  $10''$ vicinity of the pulsar, as a prior, while the color excess was fixed at $E(B-V)=0.22$.
The corresponding ratio $N_{\rm H,21}/A_V = 1.76^{+0.12}_{-0.10}$, at $A_{\rm V}=3.1 E(B-V)$,  
is in excellent agreement with the 
correlation $N_{\rm H,21}/A_V =1.79\pm0.03$  
found by \citet{Predehl_Schmitt1995}, 
although this ratio is significantly lower than $2.87\pm 0.12$ suggested by \citet{Foight2016}.
We started from fitting the absorbed brPL + BB model. 
The best fit  
is shown in the left panel of Figure~\ref{fig:opt-xrays_fit}.  
The BB temperature, $kT_{\rm BB} = 51\pm 1$ eV, is somewhat lower than those obtained from the
X-ray-only fits, 58--65 eV \citep{Karpova2014,Marelli2014,Auchettl2015}, while it is compatible with the (rather uncertain) temperature range of $\sim 40$--50 eV estimated above for the optical + FUV data.
At the same time, the emitting sphere radius $R=8.6^{+1.1}_{-1.0}$ km at the distance $d=270$ pc is 
larger than $R=3$--5 km (recalculated for the same distance) from X-ray only fits.
The break energy of the brPL component, $E_{\rm brk}\approx 0.5$ keV, is in the soft X-ray range, where the BB component dominates. The brPL slopes, $\alpha_{\rm opt} = -0.04 \pm 0.04$ 
at $E<E_{\rm brk}$ and $\alpha_{\rm X} = -1.60 \pm 0.02$ 
at $E>E_{\rm brk}$, 
correspond to the optical-FUV and X-ray parts of the pulsar's nonthermal spectrum. The brPL slope 
in X-rays
is within the range of $-1.7 <\alpha_X <-1.6$ 
found from the X-ray-only BB+PL fits, while $\alpha_{\rm opt}$ 
is at the upper 
end of the very broad interval of the PL slopes allowed by PL+BB fit of the optical-FUV spectrum (see Figure~\ref{fig:bestfit_conf}). The brPL normalization $K_{\rm PL} = 3.0^{+0.6}_{-0.4} \times 10^{-4}$ photons cm$^{-2}$ s$^{-1}$ keV$^{-1}$ at 1 keV corresponds to extrapolation of the low-energy branch.
%

With the reduced chi-square $\chi_{\rm red}^2 = 625/560=1.116$ and reasonable fitting parameters,
the fit with the brPL + BB model looks acceptable. However, 
the observed pulsations in soft (mostly thermal) X-rays with a pulsed fraction of $\sim 35\%$--40\% \citep{Marelli2014} mean that the temperature is not uniformly distributed over the NS surface, i.e., the brPL +BB model does not describe the phase-resolved X-ray data. Therefore, we considered a model with two thermal (BB) components: a cold BB component from the bulk of the NS surface and a hot BB component from small areas of the surface, presumably heated by the relativistic particles precipitating from the pulsar magnetosphere. 
The best fit gives $\chi^{2}_{\rm red} = 590/558 = 1.057$ (versus 1.116 for the BB + brPL model), which means that the second thermal component is needed with probability of 99.99999 \%, according to the F-test. 
 
In the brPL + BB +BB fit the 
fitting parameters for the colder BB component,
$kT_{\rm cold} = 41.9^{+1.4}_{-1.6}$ eV and $R_{\rm cold} = (14.7 \pm 1.1)$ km at $d = 270$ pc, are
close to the BB parameters obtained from the BB + brPL
fit. The brPL slopes, $\alpha_{\rm opt}=-0.17^{+0.05}_{-0.04}$ and $\alpha_X = -1.66 \pm 0.04$, also do not show a substantial change. The hot BB component is characterized by $kT_{\rm hot}=(83 \pm 5)$ eV and $R_{\rm hot}=(1.2 \pm 0.3)$ km at $d=270$ pc.

\begin{figure*}[!t]
\center{\includegraphics[width=0.9\linewidth]{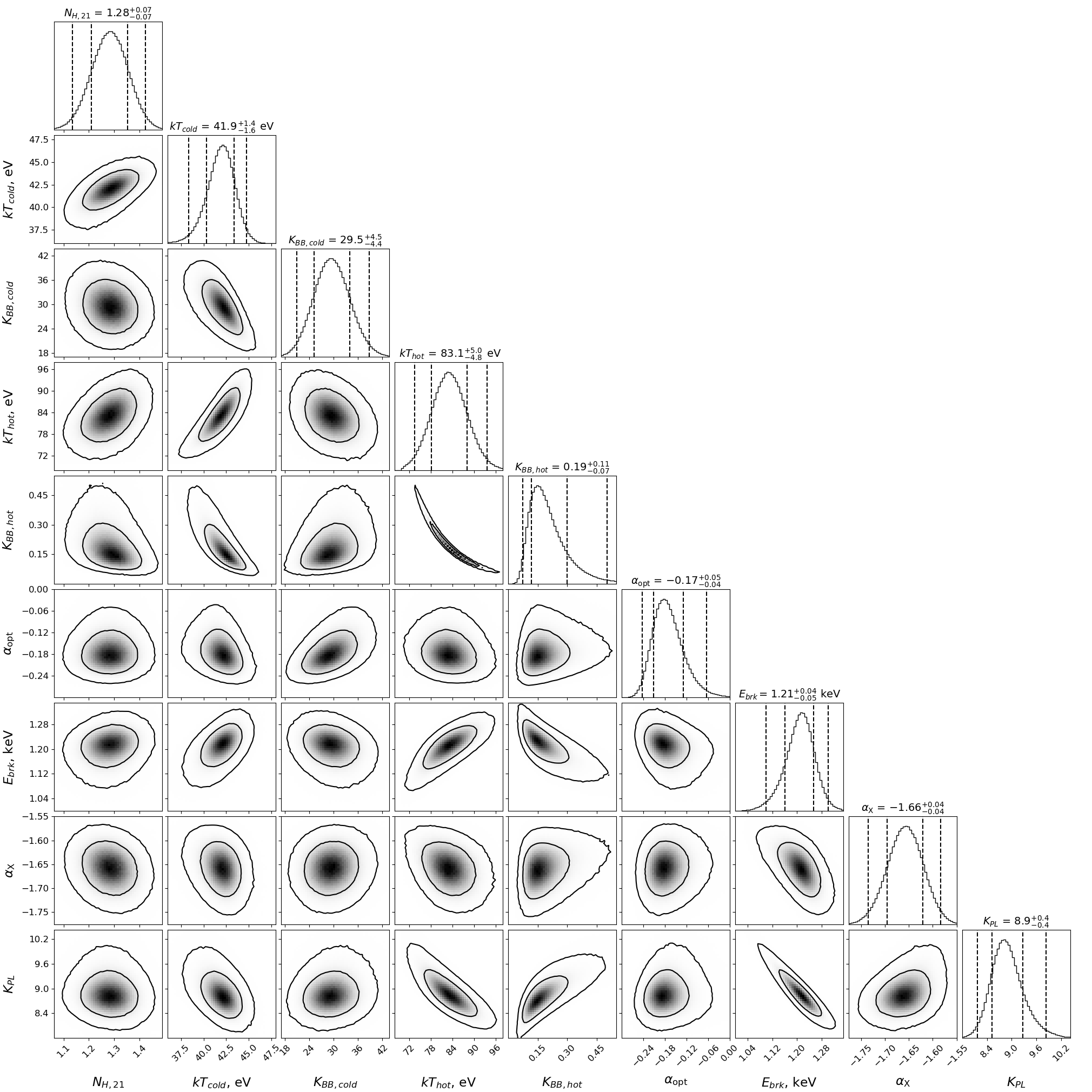}}
\caption{
Posterior probability distributions for pairs of the fitting parameters for the absorbed BB + BB + 
brPL fit of the optical-FUV + X-ray data with 68\% and 95\% percentile contours. 
Marginalized 1D parameter distribution are shown with the median parameter values and 68\% credible intervals. The vertical dashed lines correspond to 68\% and 95\% percentiles of the distributions.
The BB normalization is defined as $K_{\rm BB} = (R/10\,{\rm km})^2/(d/1\,{\rm kpc})^2$, and the brPL normalization 
$K_{\rm PL}$ is the photon flux 
of the low-energy component 
in units of $10^{-5}$ photons cm$^{-2}$ s$^{-1}$ keV$^{-1}$ at 1 keV (see the 
text for all the notations).
\label{fig:triangle_optxrays_bbbb+bknpl}}
\end{figure*}

The best-fit spectrum for the BB + BB + brPL model 
is shown in the right panel of Figure~\ref{fig:opt-xrays_fit},
while the 1D and 2D posterior probability distributions with median values and 68\% 
limits of the model parameters are presented in
Figure~\ref{fig:triangle_optxrays_bbbb+bknpl}. 
The BB normalization in Figure~\ref{fig:triangle_optxrays_bbbb+bknpl} is defined as $K_{\rm BB} = R_{10}^2/d_{\rm kpc}^2$, where $R_{10}$ is the radius of equivalent emitting sphere in units of 10 km (for a distant observer), and $d_{\rm kpc}$ is the distance in units of 1 kpc.
Thus, 
$K_{\rm BB, cold}=29.5^{+4.5}_{-4.4}$ 
corresponds to $R_{\rm cold}=14.7 \pm 1.1$ km 
at $d_{\rm kpc}=0.27$.
We see from the right panel of Figure~\ref{fig:opt-xrays_fit} that the hot BB component dominates the spectrum in a rather narrow energy range, about 0.5--0.7 keV.
Thus, adding the hot BB component does not significantly affect the overall fit, and it does not resolve the problem of the high pulsed fraction at $E\lesssim 0.5$ keV, where the (cold) BB component dominates. More complicated (likely, multi-temperature) models for the thermal emission could help, but they require a phase-resolved spectral analysis, which is beyond the scope of this work.

\begin{figure}[ht]
\center{\includegraphics[width=1.0\linewidth]{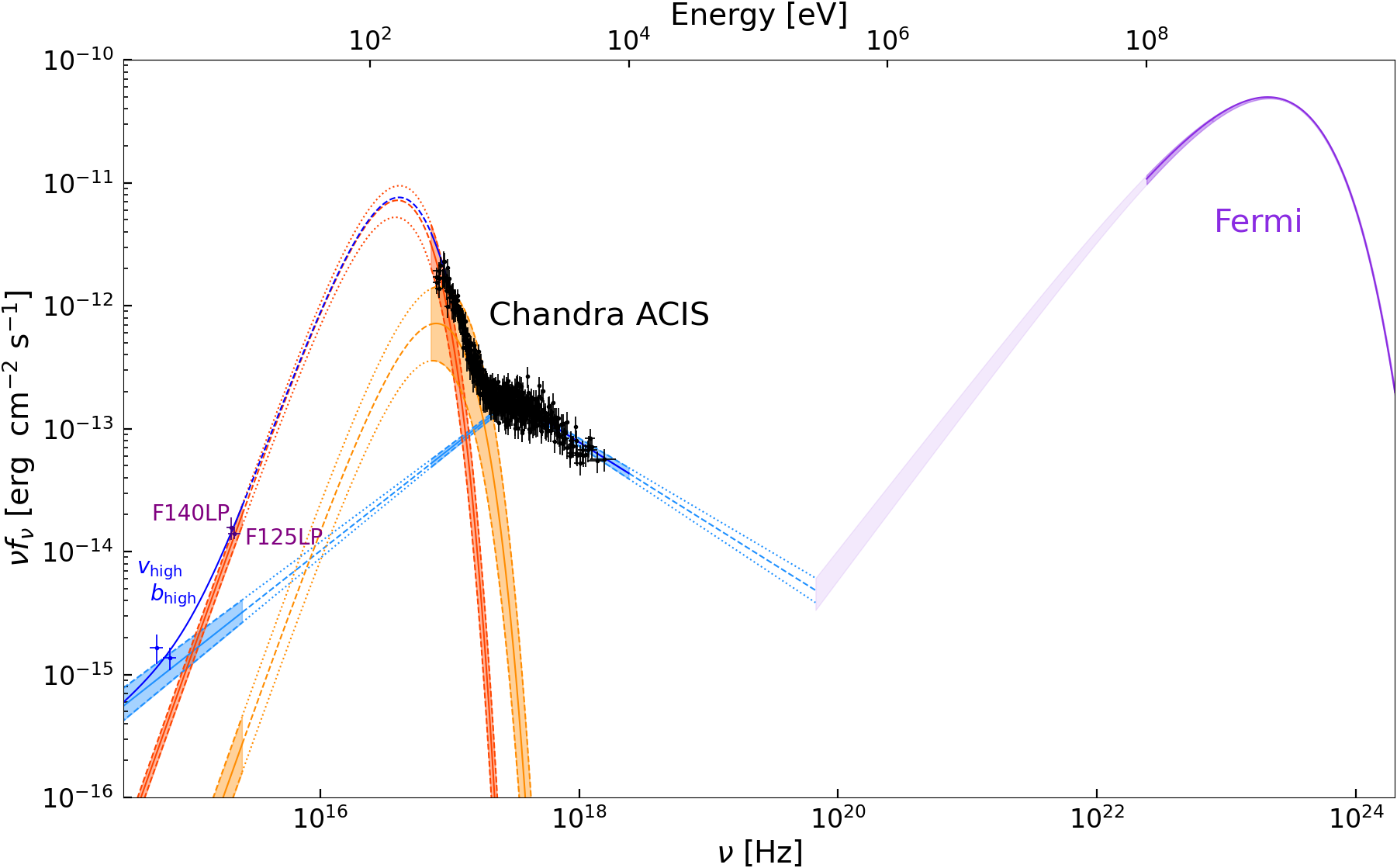}}
\caption{
SED for the unabsorbed optical through $\gamma$-ray emission from J1741.
The cold BB, hot BB and broken PL model components are shown by red-orange, orange and blue lines, respectively, the corresponding shaded areas show the 1$\sigma$ uncertainties of the model components in the ranges where the pulsar was observed. The purple lines and shaded area show the $\gamma$-ray model and its 1$\sigma$ uncertainties, whereas the light-purple shaded area 
shows the continuation of the $\gamma$-ray model toward lower energies. The blue solid line shows the best-fit model to the optical-UV and X-ray data. 
\label{fig:opt-gamma_nuF_nu}}
\end{figure}

\begin{figure}[ht]
\center{\includegraphics[width=1.0\linewidth]{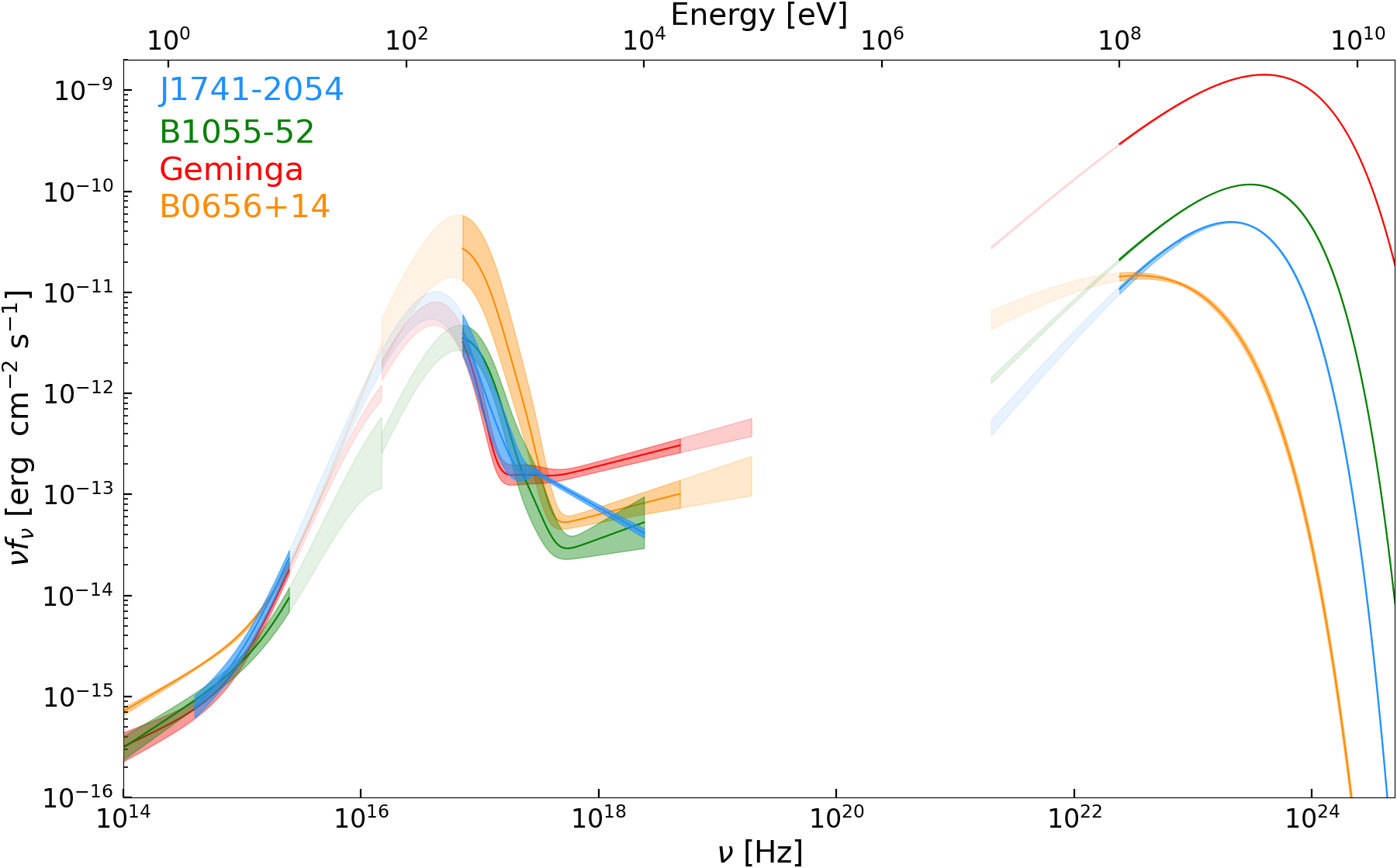}}
\caption{ 
SED for the unabsorbed IR through $\gamma$-ray emission from J1741 and other well-studied middle-aged pulsars. The IR through X-ray fit parameters for B1055-52, Geminga and B0656+14 are taken from \citet{Posselt2023}, \citet{Kargaltsev2005}, \citet{Mori2014}, \citet{Durant2011} and \citet{Arumugasamy2018}. The $\gamma$-ray fit parameters are taken from {\em Fermi} LAT 14-Year Point Source Catalog (4FGL-DR4) (\citet{Abdollahi2020}, \citet{Ballet2023}).
\label{fig:opt-gamma_nuF_nu_4}}
\end{figure}
\subsection{Bow Shock Nebula} 
\label{BSN}
Since the F125LP and F140LP filters
are rather broad, and the passband of F140LP 
is within the F125LP passband (see Figure~\ref{fig:filters}), 
we have very little information on the FUV BSN spectrum.
We do not even know whether the 
spectrum is dominated by spectral lines (like in the optical range) or by a continuum, or lines and continuum are similarly important. The only observational parameter available  that may constrain the spectral shape for a BSN element is the ratio of count rates (or average brightnesses) in the two FUV filters,
$\eta \equiv C_{\rm F125LP}/C_{\rm F140LP}$.
Table~\ref{table:C125140} shows this ratio for the six BSN regions.
The differences of the count rate ratios in different BSN regions are within their statistical uncertainties, with the weighted mean value $\overline{\eta} = 1.58 \pm 0.06$ 
for the non-overlapping elements (first five lines in Table~\ref{table:C125140}).

If the spectrum of a BSN element is dominated by a continuum, which can be approximated by a PL in the FUV range 
($f_\nu = f_{\nu,0} \nu^{\alpha}$ for the unabsorbed spectrum),
then we can, following \cite{Rangelov2016}, use the measured $\eta$ values to
estimate, with the aid of {\tt stsynphot}
\footnote{\url{https://stsynphot.readthedocs.io/en/latest/}.
}, 
the range of allowed spectral slopes $\alpha$,
at fixed values of $E(B-V)$. Then, for given $\alpha$ and extinction, we can calculate the normalization $f_{\nu,0}$
which gives the measured count rates. The calculated values of $\alpha$ and $f_{\nu,0}$ allow one to
calculate the flux $F = \int f_\nu\,d\nu$ in a range of FUV frequencies (or wavelengths) and estimate the FUV ``isotropic luminosity'', 
$L \equiv 4\pi d^2 F$.
Table~\ref{table:C125140} provides the ranges of $\alpha$ and flux intensity 
normalizations, as well as fluxes and luminosities  in the 1250--1850 \AA\ band. We see that the slopes are rather uncertain, particularly for small regions. 
The weighted mean slope for the five non-overlapping elements is $\overline{\alpha} = -0.2 \pm 0.7$.

If 
spectral line(s) dominate the spectrum, then we need to know the number of the lines and their central wavelengths  to estimate
fluxes and luminosities of the BSN elements. The flux density of a line $k$ with a width much narrower than the filter width can be approximated as $f_{\nu,k} = F_k\,\delta(\nu - \nu_k)$, where $F_k$ is the flux in the line, $\nu_k = c/\lambda_k$, $\lambda_k$  is the line's central wavelength. The contribution of such a line in the count rate of the $i$-th filter is 
$C_i^{(k)} \propto F_k T_i(\lambda_k)$, and the count rate ratio is
\begin{equation}
    \eta= \left[\sum_k F_k T_{125}(\lambda_k)\right]\left[\sum_k F_k T_{140}(\lambda_k)\right]^{-1}\,,
\end{equation}
where $T_{125}(\lambda_k)$ and $T_{140}(\lambda_k)$ are the throughputs in the F125LP and F140LP filters at $\lambda=\lambda_k$. 
In principle, the count rate ratio $\eta$ could be provided by one spectral line centered at the wavelength where the ratio $T_{125}/T_{140}= \eta$. For $\eta=1.62\pm 0.06$ 
(the Half region), 
the central wavelength of such a hypothetical line is  1368--1369 \AA, 
and the mean 
line intensity $I = F/A_s = 2\times 10^{-16}$ erg cm$^{-2}$ s$^{-1}$ arcsec$^{-2}$ 
provides the surface brightnesses 
${\cal B}_s=0.045$ and 0.028 
counts s$^{-1}$ arcsec$^{-2}$ in the F125LP and F140LP filters, respectively. 
This corresponds to the observed flux $F_{-14}\equiv F/(10^{-14}\,{\rm erg}\,{\rm cm}^{-2}\,{\rm s}^{-1})=2.9$
from the 
Half region ($F_{-14} \approx 5.8$
from the full, nearly elliptical BSN). However, as there are no strong lines at these wavelengths, 
there should be at least one line in the F140LP passband and one line in the 1230--1350 \AA\ range. Without knowing the line wavelengths, we cannot estimate the energy flux and intensity from the two count rates available. We can, however, consider the above-estimated flux in the hypothetical line as close to a lower limit. 

The observed H$\alpha$ energy
flux from the 
full J1741 BSN is 
$F_{{\rm H}\alpha,-14}^{\rm obs} \approx 2.3$
\citep{Brownsberger2014}, which corresponds to the unabsorbed flux $F_{{\rm H}\alpha,-14} \approx 3.8$.
Although we do not know the true FUV spectrum of the BSN, the above estimates of the FUV flux demonstrate that it exceeds the H$\alpha$ flux. 
For instance, assuming that the 
spectrum can be described by a PL model, we estimated the
unabsorbed 1250--1850 \AA\ flux $F_{\rm FUV,-14} \approx 27$ for the Half region ($\approx 54$ for
the full BSN), at $E(B-V)=0.22$, which 
corresponds to the FUV luminosity $L_{\rm FUV}\approx 4.7 \times 10^{30} d^{2}_{270}$ erg s$^{-1}$.
The unabsorbed FUV flux and luminosity exceed those in the H$\alpha$ line by
a factor of 14.
\begin{table}[ht]
\setlength{\tabcolsep}{0.5em}
\renewcommand{\arraystretch}{1.25}
\caption{Properties of BSN regions}
\label{table:C125140}      
\centering          
\begin{tabular}{l c c D{,}{\,\pm\,}{-1} c c}     
\hline\hline                
region & $\eta$ & $\alpha$ & \multicolumn{1}{c}{$I_{\nu,0}$} & $F_{-14}$ & $L_{29}$ \\
\hline                    
\rule{0pt}{2.5ex}Apex & $1.32\pm0.21$ & $-5.2^{+3.3}_{-6.9}$ & \multicolumn{1}{c}{$1060^{+90}_{-300}$} & $1.3$ & 1.1\\
Front & $1.58\pm0.10$ & $-1.3^{+1.0}_{-1.3}$ & \multicolumn{1}{c}{$610^{+20}_{-30}$} & $9.7$ & 8.4 \\
Back & $1.54\pm0.11$ & $-1.8^{+1.3}_{-1.6}$ & \multicolumn{1}{c}{$159^{+5}_{-7}$} & $7.6$ & 6.6\\
Blob & $1.33\pm0.16$ & $-5.1^{+2.6}_{-4.5}$ & \multicolumn{1}{c}{$420^{+20}_{-70}$} & $3.0$ & $2.6$\\
Interior & $2.01 \pm 0.15$ & $+2.7^{+1.1}_{-1.2}$ &  130,10 & $11$ & $9.7$\\
Half & $1.62\pm0.06$ & $-1.0 \pm 0.7$ & 243,6 & $27$ & 24\\ 
\hline 
\end{tabular}
\tablefoot{The second column presents the count rate ratios, $\eta=C_{\rm F125LP}/C_{\rm F140LP}$. The third through sixth columns 
provide properties of the BSN regions assuming an unabsorbed PL spectrum $f_\nu = f_{\nu,0} (\nu/\nu_0)^{\alpha}$ for each of them;
$I_{\nu,0} = f_{\nu,0}/A_s$ is the specific intensity normalization 
at $\lambda=1500$ \AA, 
in units of nJy arcsec$^{-2}$; $F_{-14}$ is the unabsorbed flux estimate in the 1250--1850 \AA\ FUV band, in units of $10^{-14}$ erg cm$^{-2}$ s$^{-1}$ , and $L_{29}$ is the FUV luminosity 
in units of $10^{29}$ erg s$^{-1}$. The color 
excess $E(B-V)=0.22$ is assumed.}
\end{table}


\section{Discussion}

The HST FUV observation presented here allowed us to to detect
with high confidence the UV counterpart, and confirm the optical counterpart, of the middle-aged J1741 pulsar. In addition, we detected the 
FUV BSN created by this pulsar. 
Below we discuss the inferred properties of the J1741 pulsar and its BSN and compare them with the properties of similar objects.

\subsection{The pulsar}
Our analysis of the HST FUV and archival VLT optical data on the J1741 pulsar has shown that most likely the optical-UV emission of this pulsar consists of two components representing  nonthermal emission from 
the 
pulsar magnetosphere and thermal emission from the NS surface. It is difficult to confidently constrain the PL slope of the optical-FUV nonthermal component with the  currently available data in this range alone.  Additional observations in the near-IR and near-UV are required to reliably constrain the PL slope and temperature of the bulk of the NS surface. 

Adding X-ray archival  data to the analysis allowed us to fit the multiwavelength spectrum of the pulsar with the brPL + BB or brPL + BB + BB  models.
In these 
models the 
low-energy part of 
the nonthermal emission, from the optical to 
the break energy of about 0.5 or 1.2 keV, is 
described by 
a hard PL with photon index $\Gamma_{\rm opt} \approx 1.0$--1.2 ($-\alpha_{\rm opt}\approx 0$--0.2),
while  one or two BB thermal components
plus a soft PL component with $\Gamma_X \approx 2.6$--2.7 ($-\alpha_X \approx 1.6$--1.7) are needed 
to describe the pulsar spectrum in X-rays up to 10 keV. 
According to these fits, the temperature of the bulk of the NS surface, $kT_{\rm cold}\approx 40$--52 eV, is significantly 
lower than $60\pm2$ eV value obtained from the X-ray data alone (see Section \ref{intro}).
As can be seen from Figure 5 of \citet{Karpova2014}, 
the lower temperature makes the position of J1741 in the NS temperature-age plane  fully consistent with the predictions of the standard  NS cooling models. The 
temperature of the hot BB component in the brPL + BB + BB fit, $kT_{\rm hot} \sim 80$ eV, could be ascribed to a small part of the NS surface, possibly heated by relativistic particles precipitating from the NS magnetosphere or by the anisotropic heat transfer from the NS interiors.

Although the spectral fits are formally acceptable, we have to note that they are hardly consistent with the high pulsed fraction of the (cold) thermal component \citep{Marelli2014}, at least for the locally semi-isotropic BB emission from NS surface elements. 
This suggests that the BB + BB model is too simplistic to describe the thermal emission from J1741. 
In particular, it seems quite plausible that, instead of two 
surface regions with different temperatures and emitting areas, there 
is a smooth temperature distribution over the NS surface caused by the anisotropic heat transfer in a strong magnetic field \citep{Greenstein1983}. 
For instance, if an NS possesses a dipole magnetic field near the surface, its magnetic poles are warmer than the magnetic equator, even without external heating of the poles by relativistic particles. According to \citet{Yakovlev2021a,Yakovlev2021b}, the BB + BB model can mimic thermal spectrum from an NS with such a temperature distribution, 
under the assumption that the surface elements emit semi-isotropic BB with the local temperatures.
These models, however, do not reproduce the observed high pulsed fraction of thermal emission.
The high pulsed fraction can be reached if there is a strong toroidal component of the magnetic field 
under the NS surface \citep{Geppert2006,Igoshev2021}. 

A higher model pulsed fraction can also be obtained if one takes into account the anisotropy of local emission from a magnetized NS surface.  
Bombardment of NS polar regions 
by relativistic electrons or positrons not only additionally heats these regions, but it can also produce protons and $\alpha$-particles in spallation reactions, i.e., a light-element plasma atmosphere can be formed above the polar caps  
\citep{Ho2007}.
The highly anisotropic photon emission emergent from such an atmosphere 
\citep{Pavlov1994} can explain the high pulsed fraction of the hot thermal component. 

The cold thermal component is expected to come 
from colder surface regions that can be formed by a condensed (liquid-like or solid-state) matter in a strong magnetic field (e.g., \citealt{Medin2007}). The spectrum of its emission is not strongly different from the BB spectrum  depressed by a constant factor \citep{Adelsberg2005,Peerez-Azorin2005}, but the angular distribution of 
its specific intensity  should be anisotropic. We are not aware of detailed investigation of pulsed emission from a rotating NS with a condensed surface (but see \citealt{Peerez-Azorin2005}).   
Consideration of more complicated surface emission models is, however, beyond the scope of this paper.

Since J1741 is a $\gamma$-ray pulsar, 
we can compare the optical through X-ray nonthermal spectrum with the $\gamma$-ray spectrum. Figure~\ref{fig:opt-gamma_nuF_nu} shows the multiwavelength J1741's spectral energy distribution (SED) from optical to $\gamma$-ray photon energies, using 
the {\em Fermi} LAT results 
\citep{Abdollahi2020,Ballet2023}. We see that the nonthermal spectrum has at least two spectral breaks, at $\sim 1$ keV and $\sim 200$ keV.
It suggests that the nonthermal emission is not produced by a single physical mechanism, and it likely comes from distinct sites.
We, however, cannot exclude that more observations in the IR-optical-UV range and more realistic models of thermal emission can significantly change the optical and X-ray PL slopes, so that, for instance, the optical PL may become softer while the X-ray PL may become harder (and maybe even dominated by the hot thermal emission in the {\em Chandra/XMM-Newton} range). 
However, it seems impossible to describe the 
nonthermal 1\,eV -- 100\,MeV spectrum by a single PL.

With a characteristic age of 386 kyr and distance $\sim 270$ pc, the J1741 pulsar belongs to a small group of nearby, middle-aged pulsars detected in the FUV, which include Geminga, B0656+14, and B1055-52 (B0656 and B1055 hereafter), sometimes called the Three Musketeers \citep{Becker1997}.
The multiwavelength 
spectrum of J1741 is compared with those of the Three Musketeers in Figure~\ref{fig:opt-gamma_nuF_nu_4}, while the spectral parameters of all the four pulsars are compiled in  Table~\ref{table:4PSRs2}.
Note that the spectral models used for fitting were not exactly the same for the four pulsars. The parameters of J1741 were obtained from a joint optical through X-ray fit with the brPL + 2BB model, while the X-ray and optical-UV spectra of 
B1055 and Geminga were fitted separately, 
with the PL + BB + BB and PL + BB models, respectively, and both models were applied to B0656 (with addition of an absorption feature at 0.5--0.6 keV). Therefore, in Table~\ref{table:4PSRs2}, the positions of the spectral break ($E_{\rm brk}$) in the nonthermal emission component  are given  only for J1741 and B0656. 

We see from Figure~\ref{fig:opt-gamma_nuF_nu_4} that the phase-integrated multiwavelength spectra of the four middle-aged pulsars are qualitatively similar. All of them have prominent ``thermal humps'', seen above the nonthermal spectrum from $\sim$5 eV to $\sim$2 keV. 
For all the four pulsars,  
these humps can be 
described by combinations of cold and hot BB components. 
The main 
contribution to the thermal humps comes from the cold components, with temperatures $kT_{\rm cold}\sim 40$--70 eV, 
radii of equivalent emitting spheres $R_{\rm cold}\sim 5$--15 km, and bolometric luminosities $L_{\rm cold}\sim (0.6$--$3.6)\times 10^{32}$ erg s$^{-1}$. The Rayleigh-Jeans tails of the cold thermal component have been detected in the FUV ($E\sim 6$--10 eV) for all the four pulsars. 
The FUV fluxes were included in the brPL + 2BB fits of the spectra of two of the four pulsars, B0656 and J1741, which resulted in somewhat different $kT_{\rm cold}$ compared to X-ray-only fits.

We see from Table~\ref{table:4PSRs2}, that $T_{\rm cold}
\sim 42$ eV, $R_{\rm cold}\sim 15$ km (for the estimated distance of 270\,pc), and $L_{\rm cold}\sim 0.9\times 10^{32}$ erg s$^{-1}$, of J1741 are within the temperature and luminosity ranges for the Three Musketeers.
As the radii of J1741, B0656 and Geminga are comparable with an expected NS radius, one can assume that the cold thermal emission comes from a substantial part of the NS surface and is likely powered by the heat 
supplied by  hotter NS interiors. 
The temperatures and luminosities do not show a clear decrease with increasing the pulsars' characteristic ages, 
but we should not forget that true ages may differ from characteristic ages, and the distances to three of the four pulsars are not well known, which affects the radius and, particularly, luminosity estimates. 

For the cases where the optical-UV and X-ray spectra were fitted separately, the temperatures $T_{\rm FUV}$ in Table~\ref{table:4PSRs2} are brightness temperatures estimated from the measured FUV flux densities at an assumed (fixed) NS radius of 15 km and the best distance estimates available. Since $f_\nu \propto (R/d)^2 T$ in the Rayleigh-Jeans regime, and the best-fit radii $R_{\rm cold}$ are smaller than assumed $R_{\rm FUV}=15$ km for the Three Musketeers, it is not surprising that $T_{\rm FUV}$ is lower than $T_{\rm cold}$.

Additional hot BB components,
with close temperatures $kT_{\rm hot}\sim 80$--200 eV,
are seen in the X-ray emission of all the four middle-aged pulsars.
The effective radii of J1741 and B0656, are close to each other, $R_{\rm hot}\sim 1$ km, but they are significantly larger than $R_{\rm hot}\sim 200$ m and 45 m for B1055 and Geminga, respectively. 
The corresponding bolometric luminosities, $L_{\rm hot} \sim (0.4$--$30)\times 10^{30}$ erg s$^{-1}$,
also show a strong scatter, with the Geminga's luminosity being an obvious outlier. The luminosities 
do not show  correlation with the spindown power, as would be expected for heating by magnetospheric particles, but there is hint of positive correlation with $L_{\rm cold}$ (at least if we exclude the outlier Geminga). 
This suggests that the 
anisotropic heat transfer from the NS interior indeed plays a role in heating the polar regions.

The ratio 
of the 
hot BB and cold BB temperatures, 
obtained from the phase-averaged fits, $T_{\rm hot}/T_{\rm cold} \approx 2.0$ for J1741, is close to $2.4$ for B1055
and $1.9$ for B0656,
%
%
%
but it is substantially lower than 
 $4.4$ 
for Geminga.  
According to \citet{Yakovlev2021a}, 
the anisotropic heat transfer alone cannot explain such high temperature ratios for a dipole magnetic field (typical values do not exceed 1.2 in those calculations). This means that either there is an additional heating mechanism of the hot regions or the magnetic field is different from a centered dipole (e.g., there is a strong toroidal component under the NS surface). 
Different temperature ratios of the four pulsars can come from a combination of  differences in the bulk temperatures distributions, hot spot temperatures (influenced by different magnetic field topology and strengths), and pulsar geometries (rotation-magnetic axes inclinations and viewing angle)  
as well as possible occultations of the surface by plasma-filled patches in the magnetopshere
\citep{Kargaltsev2012}.
Different 
pulsed fractions in the thermal components, from $\sim$36\,\% for 
J1741 \citep{Marelli2014} to $\sim$65\,\% for 
B1055 \citep{Vahdat2024}
are additional observational manifestations of these effects. 
Nevertheless, it is remarkable that the temperature ratios for the three pulsars are in a relatively narrow range around  $2$ even if we take into account their variations with the pulsar rotation phase (e.g., $T_{\rm hot}/T_{\rm cold}$ varies from $1.9 \pm 0.2$ to  $2.7 \pm 0.2$ for 
B1055 and from $1.8 \pm 0.2$ to $2.2 \pm 0.3$ for B0656 \citep{Vahdat2024,Arumugasamy2018}).

The distance-independent ratios of BB radii and bolometric luminosities for the cold and hot components
are also similar for 
J1741, B1055 and B0656 ($R_{\rm hot}/R_{\rm cold} \sim L_{\rm hot}/L_{\rm cold} \sim 0.04$--0.10), 
but they are significantly smaller for Geminga (0.004 and 0.006 for the radii and luminosties ratios, respectively).  
These ratios are much smaller than calculated by 
\citealt{Yakovlev2021a} for the anisotropic heat transfer in a dipole magnetic field, 
which can be considered as an additional argument against such a simplified picture. 
\citet{Rigoselli2022} have noticed that six rotation powered pulsars of different ages whose thermal spectra are  well described by the double BB model, including the Three Musketeers except of Geminga,    occupy a remarkably compact region  in the   $(T_{\rm hot}/T_{\rm cold})$--$(R_{\rm hot}/R_{\rm cold})$ plane.  This may imply similar surface temperature maps for these pulsars independently of their 
ages.
J1741 obviously joins to this group.

All the four pulsars show qualitatively similar nonthermal spectra in the optical to X-ray range (Figure~\ref{fig:opt-gamma_nuF_nu_4}). If fitted by a PL model $f_\nu\propto \nu^\alpha = \nu^{1-\Gamma}$ separately in the two ranges, the slope $\alpha$ is more negative in X-rays than in the optical (e.g., $\alpha_{\rm opt} -\alpha_{\rm X} = \Gamma_X -\Gamma_{\rm opt}\sim 0.2$--0.4 for the Three Musketeers). 
It means that the nonthermal spectrum can be fitted with a broken PL 
in the optical through X-ray range, with a break energy $E_{\rm br}$ between UV and hard X-rays. 
The break energy is rather uncertain for the Three Musketeers because the slope differences are small (see the example of B0656 in the last column of Table~\ref{table:4PSRs2}), but it is more certain for J1741, for which $\alpha_{\rm opt} -\alpha_{\rm X} = 1.49 \pm 0.08$. This large value is due to the
unusually steep X-ray PL slope, $\Gamma_X \approx 2.7$ versus 1.3--1.8 for the Three Musketeers, which is particularly well seen in Figure~\ref{fig:opt-gamma_nuF_nu_4}.
%
The large $\Gamma_X$ is an outlier not only in the small sample of four pulsars but also among a large sample of pulsars with similar parameters, such as 
spindown power or characteristic age (see Figures 3 and 4 in \citet{Chang2023}).

Similar to the other spectral parameters, the X-ray PL slope (photon index) varies with rotational phase.
The relatively narrow 
photon index range in the phase-resolved spectra of J1741 ($\Gamma_X\sim 2.3$--2.8; \citealt{Marelli2014}) slightly overlaps with the respective 
ranges of B1055 ($\Gamma_X\sim 1.5$--2.5; \citealt{Vahdat2024}), Geminga ($\Gamma_x\sim 1.6$--2.2; \citealt{Mori2014}), and B0656 ($\Gamma_X\sim 1.5$--2.3; \citealt{Arumugasamy2018}). 
A possible explanation for the J1741's 
offset $\Gamma_X$ values and the relatively large ranges spanned by the photon indices of the Three Musketeers  could come from different pulsar geometries 
that allow one to 'see' different parts of nonthermal emission regions around the pulsar. 
There is also the possibility that the J1741 spectrum has another phase-dependent thermal component, invisible with the current count statistics. However, our attempts to describe the high energy spectral tail by an extra BB component instead of the PL  led to unacceptable fits.   
A deeper, phase-resolved X-ray study of J1741 could clarify this.

We see from Figure~\ref{fig:opt-gamma_nuF_nu_4} that the nonthermal X-ray  
spectra do not connect smoothly 
with the $\gamma$-ray spectra for all the four pulsars. It could indicate that the X-rays and $\gamma$-rays
come from different sites in the magnetosphere and/or are produced by different emission mechanisms. This inference is less certain for J1741 and B1055, which have not been observed in hard X-rays (above $\sim$10 keV).

We also see that the SEDs of the four pulsars
reach their maxima in $\gamma$-rays, at energies from 
$\sim 100$ MeV for B0656 to $\sim 2$ GeV for Geminga.
This implies that the most efficient
conversion of the pulsar rotation energy
to radiation occurs in  $\gamma$-rays.
The efficiency in a spectral band $b$ is usually defined as $\eta_{\rm b}=L_{\rm b}/\dot{E}$,  
where $L_{\rm b}$ is the (distance-dependent) luminosity in the band. Comparing the 
efficiencies  $\eta_{\gamma}$ 
estimated for the 0.1--100 GeV band (Table~\ref{table:4PSRs2}), we see that Geminga is 
the most efficient emitter in this band, B1055 and J1741 are intermediate ones, while B0656 has the lowest  efficiency, in accord with the SEDs in Figure~\ref{fig:opt-gamma_nuF_nu_4}.     
In the optical-FUV (1--10 eV) and X-ray (1--10 keV) bands (where the efficiencies are lower than in $\gamma$-rays by about 5 and 3 
orders of magnitude, respectively),
J1741 
seems to be the most efficient emitter  among the four pulsars.    
 Besides the obvious uncertainties due to uncertain distances, we have to note, however, 
 that the nonthermal luminosity estimate of J1741 in the 1--10 eV band is based on the optical + X-rays 
 fit with the brPL + 2BB model involving only four data-points in this band. 
 New observations in the near-IR and near-UV are needed to add more data-points and confirm the high 
 efficiency of J1741 in the optical-FUV band. 


\begin{table*}
\renewcommand{\arraystretch}{1.25}
\caption{Optical-FUV and X-ray properties of J1741 and the Three Musketeers}             
\label{table:4PSRs2}      
\centering          
\begin{tabular}{llllll}     
\hline\hline                
\rule{0pt}{2.1ex}PSR &  J1741-2054 & B1055-52 & Geminga & \multicolumn{2}{c}{B0656+14}\\
\hline 
$\tau$, kyr & 386 & 535 & 342 & \multicolumn{2}{c}{111} \\
$\dot{E}_{34}$ & 0.95 & 3.0 & 3.2 & \multicolumn{2}{c}{3.8}\\
$B_{12}$ & 2.7 & 1.1 & 1.6 & \multicolumn{2}{c}{4.7}\\
$d$, pc & $\sim 270$ & $\sim$350 & $250^{+230}_{-80}$ & \multicolumn{2}{c}{$290\pm30$}\\
$N_{\rm H, 20}$ & $13 \pm 1$ & $1.4^{+0.3}_{-0.2}$ & $1.5 \pm 0.3$ & $3.0^{+0.7}_{-0.9}$ & $3.1^{+0.4}_{-0.2}$\\ 
$kT_{\rm cold}$, eV &  $42 \pm 2$ & $70\pm2$ & $44\pm1$ & $64 \pm 4$ & $68 \pm 1$\\ 
$R_{\rm cold}$, km & $14.7 \pm 1.1$ & $4.7^{+0.5}_{-0.4}$ & $11.4 \pm 1.1$ & $13^{+4}_{-3}$ & $11.2^{+0.4}_{-0.5}$\\ 
$L_{\rm cold,32}$ & $0.86$ & 0.69 & 0.63 & 3.66 & 3.48\\ 
$kT_{\rm hot}$, eV &  $83 \pm 5$ & $165 \pm 12$ & $195 \pm 14$ & $123^{+6}_{-5}$ & $131^{+6}_{-3}$\\ 
$R_{\rm hot}$, m & $1200 \pm 300$ & $180^{+50}_{-30}$ & $45 \pm 7$ & $1000 \pm 200$ & $740^{+90}_{-120}$\\ 
$L_{\rm hot, 30}$ & $8.8$ & 3.1 & 0.38 & 30 & 21\\
$T_{\rm hot}/T_{\rm cold}$ & $2.0$ & 2.4 & 4.4 & 1.9 & 1.9\\
$R_{\rm hot}/R_{\rm cold}$ & $0.082$ & 0.038 & 0.004 & 0.077 & 0.066\\
$L_{\rm hot}/L_{\rm cold}$ & $0.102$ & 0.045 & 0.006& 0.082 & 0.060\\
$\alpha_{\rm X}$ & $-1.66 \pm 0.04$ & $-0.57^{+0.25}_{-0.26}$ & $-0.70 \pm 0.04$ & $-0.7 \pm 0.1$ & $-0.74^{+0.11}_{-0.12}$\\ 
$f_{\rm 0, X}$, nJy & $59 \pm 3$ & $8.1 \pm 0.3$ & $52 \pm 2$ & $17 \pm 3$ & $25^{+12}_{-7}$\\
$E_{\rm brk}$, keV & $1.21^{+0.04}_{-0.05}$ &\ldots  & \ldots & \ldots & $0.130^{+0.470}_{-0.125}$ \\ 
$\alpha_{\rm opt}$ & $-0.17^{+0.05}_{-0.04}$ & $-0.24 \pm 0.10$ & $-0.46 \pm 0.12$ & $-0.32 \pm 0.05$ & $-0.44^{+0.06}_{-0.05}$\\ 
$f_{0,\rm opt}$, nJy & $\sim 150$ & $178 \pm 25$ & $110 \pm 20$ & $340 \pm 7$ & $\sim$280\\ 
$kT_{\rm FUV}$, eV & $=kT_{\rm cold}$ & $22 \pm 6$ & $30\pm2$ & $29 \pm 2$ & $=kT_{\rm cold}$\\
$R_{\rm FUV}$, km & $=R_{\rm cold}$ & 15 & 15 & 15 & $=R_{\rm cold}$\\
$E(B - V)$ & $0.22$  & 0.03 & 0.03 & \multicolumn{2}{c}{0.03}\\
$L_{\rm 1-10\,eV, 28}$ & $2.8$ & 5.5 & 1.7 & 7.3 & 6.0 \\
$\eta_{\rm FUV,\,-6}$ & $2.9$ & 1.8 & 0.5 & 1.9 & 1.6 \\
$L_{\rm 1-10\,keV, 30}$ & $1.9$ & 1.1 & 3.1 & 1.4 & 1.1\\
$\eta_{\rm X,\,-4}$ & $2.0$ & 0.4 & 1.0 & 0.4 & 0.3 \\
$L_{\rm 0.1-100\,GeV, 33}$ & $1.0$ & 4.4 & 32 & \multicolumn{2}{c}{0.27} \\
$\eta_{\rm \gamma}$ & $0.11$ & 0.15 & 0.99 & \multicolumn{2}{c}{0.0072}\\[0.25ex]
\hline                  
\end{tabular}
\tablefoot{
The parameters in the 2nd (J1741) and 6th (B0656) columns are for joint optical through X-ray fits with the brPL + BB + BB model, while they are for separate fits in the optical (PL + BB) and X-ray (PL + BB + BB) ranges in the 3rd (B1055), 4th (Geminga) and 5th (B0656) columns. 
$L_{\rm cold, 32}$ and $L_{\rm hot, 30}$ are bolometric luminosities of the cold and hot components in units of $10^{32}$ and $10^{30}$ erg s$^{-1}$, respectively; 
$f_{0,\rm opt}$ is the optical PL normalization at $\nu_{0} = 10^{15}$ Hz;
$f_{\rm 0, X}$ is the X-ray PL normalization at $E=1$ keV; 
$kT_{\rm FUV}$ in the 3rd through 5th columns is the brightness temperature estimated from the FUV flux density at the assumed NS radius $R_{\rm FUV}=15$ km and best-fit distance;
$L_{\rm 1 - 10\,eV, 28}$, $L_{\rm 1 - 10\,keV, 30}$ 
and $L_{\rm 0.1 - 100\,GeV, 33}$ are nonthermal luminosities in 1--10 eV, 1--10 keV 
and 0.1--100 GeV ranges in units of $10^{28}$ erg s$^{-1}$, $10^{30}$ 
erg s$^{-1}$ and $10^{33}$ erg s$^{-1}$, respectively; $\eta_{\rm FUV,\,-6}$ (in units of $10^{-6}$), $\eta_{\rm X,\,-4}$ (in units of $10^{-4}$) and $\eta_{\rm \gamma}$  
are the corresponding radiation efficiencies. 
The separate X-ray fit parameters for B1055, Geminga and B0656  are taken from \citet{Posselt2023}, \citet{Mori2014} and \citet{Arumugasamy2018}, respectively, while they are from \citet{Zharikov2021} for the brPL + 2BB fit for B0656 (6th column). The FUV fit parameters are taken from \cite{Posselt2023}, 
\citet{Kargaltsev2005} and \citet{Durant2011}.
}
\end{table*}
\subsection{The bow shock nebula}

\begin{table*}
\caption{Properties of the three known FUV BSNe}             
\label{table:BSNe}      
\centering          
\begin{tabular}{cccccccccccccc}     
\hline\hline                 
\rule{0pt}{2.2ex}PSR & 
$d$ & $P$ & $\tau$ & $\dot{E}_{33}$ & $v_{\perp}$ & $E(B - V)$ & $F^{\rm H\alpha}_{-14}$ & $F^{\rm FUV}_{-14}$ & $L^{\rm FUV}_{29}$ & $\theta_{a, \perp}$ & $l_{a, \perp, \rm 15}$ & $\eta$ \\
 & pc & ms  & Myr & & km/s &  &  &  &  & $''$ & &\\[0.5ex]
\hline                    
\rule{0pt}{2.2ex}J1741$-$2054  & 270 & 414 & 0.39 & 9.5 & 140 & 0.22 & 2.3 & 10\,[$>$6] & 47\,[$>$28] & 1.1 & 4.4 & $1.6\pm0.1$\\

J0437$-$4715  & 157 & 5.76 & 1600 & 5.5 & 104 & 0.05 & 3.3 & 15--22 & 7--10 & 10 & 23 & $1.6\pm0.1$ \\
J2124$-$3358 & 410 & 4.93 & 3800 & 6.8 & 101 & 0.03 & 1.7 & 5 & 13 & 2.5 & 15 & $1.3\pm0.2$ \\
\hline                  
\end{tabular}
\tablefoot{
The second through sixth columns show the distances, periods, characteristic ages, 
spindown powers
in units of $10^{33}$ erg s$^{-1}$, and transverse velocities 
of the pulsars with detected FUV BSNe. 
The eighth and ninth columns show the observed (absorbed) fluxes 
from the entire BSNe
in H$\alpha$ 
and FUV (in 1250--1850 \AA\ range), 
in units of $10^{-14}$ erg cm$^{-2}$ s$^{-1}$. We recalculated the J0437 and J2124 BSNe FUV fluxes for 1250--1850 \AA\ range, using results, obtained by \citet{Rangelov2016,Rangelov2017}. 
The tenth column shows 
the corresponding FUV luminosities in units of $10^{29}$ erg s$^{-1}$. 
The angular distances 
between the pulsar and the forward bow shock apex and 
the corresponding projected distances in units of $10^{15}$ cm  are in the eleventh and twelfth columns, while the thirteenth column shows the ratio of count rates in the F125LP and F140LP filters.
The FUV fluxes were estimated assuming PL BSN spectra. 
The two J0437 BSN FUV flux values correspond to different background choices 
(see 
\citealt{Rangelov2016}).
The lower limits in square brackets for the J1741 were obtained assuming a line-dominated spectrum.
The FUV fluxes and luminosities 
for entire J0437 and J2124 BSNe were crudely estimated using the apex FUV fluxes
from \citet{Rangelov2016,Rangelov2017} and 
assuming similar FUV and H$\alpha$ brightness distribution for these BSNe.
}
\end{table*}

\begin{figure}[ht]
\center{\includegraphics[width=0.75\linewidth]{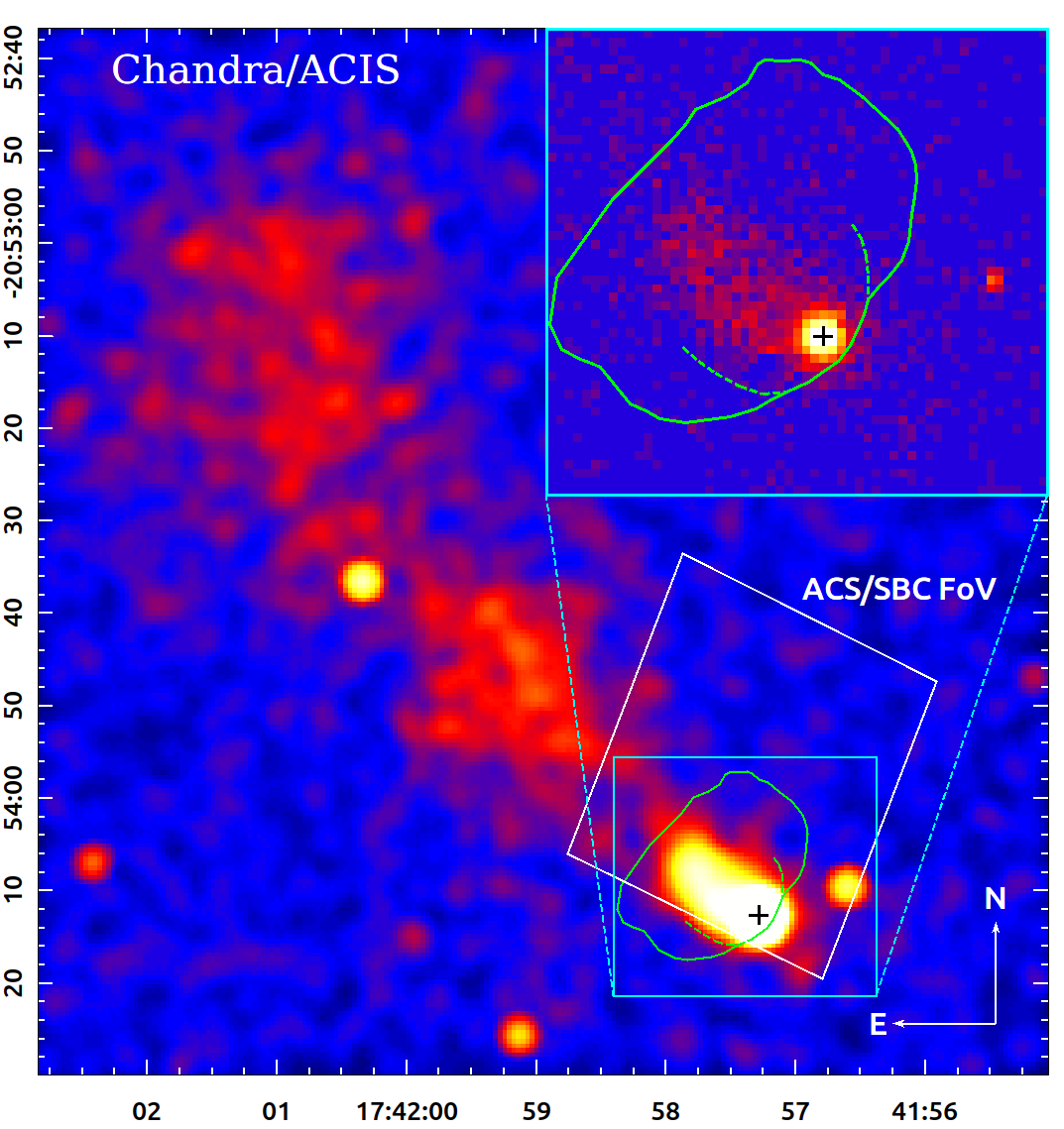}}
\caption{Comparison of the J1741 X-ray PWN and  H$\alpha$-FUV BSN shapes. The green 
contour shows the outer boundary of the H$\alpha$ BSN as imaged by SOAR (see Figure~\ref{fig:SBC_FoV}), which virtually coincides with outer boundary of the FUV BSN,
overlaid on the {\em Chandra} ACIS-S 0.7--8 keV image.
The large-scale {\em Chandra} image is smoothed with a 5 pixel Gaussian kernel, while the zoomed-in $28'' \times 26''$ image of the compact PWN in the inset
is shown without smoothing. Green 
dashed curves show continuation of the ``bulge'' into inner BSN region. The black cross shows the pulsar position. 
}
\label{fig:PWN-X}
\end{figure}

The J1741 BSN, detected in our FUV observation for the first time,
 is one of 
 only nine detected
Balmer-dominated 
BSNe, best-studied in the H$\alpha$ emission (\citealt{Romani2010}, \citealt{Brownsberger2014}). 
The paucity of Balmer-dominated BSNe indicates that most pulsars move in an ionized ISM.
The J1741's BSN is visible in the VLT $b_{\rm high}$ image due to the H$\beta$ and H$\gamma$ emission lines (\citealt{Mignani2016}; see also Figure~\ref{fig:F125LP_b-high}). 
We found no differences in
the shape and size of the FUV and H$\alpha$ BSNe, moving together with the pulsar. 
The BSN is confined within a closed shell, resembling
an ellipse 
with a flattened back, elongated perpendicular to the proper motion direction. 
In the high-resolution SOAR and HST images (Figure~\ref{fig:SBC_FoV}) we see a ``bulge''
at the sharp, bright front  of the shell, 
with a hint of continuation of the bulge boundary into the (projected) BSN interior. 
The bulge and its continuation (if confirmed with deeper high-resolution observations) can be interpreted as a forward bow shock, while the rest of the BSN could be a ``bubble'', similar to those in the famous Guitar BSN (\citealt{deVries2022}, and references therein). There is also a slight hint of a bent ``trail'' in the northeast direction, starting from the Blob region, which 
might be interpreted as
a northwest boundary of a second ``bubble''.

This interpretation of the BSN structure is supported by the comparison of the X-ray PWN and the FUV/H$\alpha$ BSN shown in Figure~\ref{fig:PWN-X}. We see that the compact component of the X-ray PWN, presumably created by the shocked pulsar wind, is seen almost up to the BSN (bulge) apex ahead of the moving pulsar, as expected (some X-ray counts ahead of the apex are due to the wings of the pulsar's PSF). However, 
this compact PWN behind the pulsar, elongated along the pulsar motion direction,
is much narrower than the BSN. 
On the other hand, the possible boundaries of the bulge continuation, shown by dashed green curves in the inset of Figure~\ref{fig:PWN-X}, are much closer to the compact X-ray PWN component, as expected from simulations of nebulae created by supersonically moving pulsars (e.g., \citealt{Bucciantini2005}).

The J1741 BSN is only the third FUV BSN 
detected, and the first BSN associated with a middle-aged pulsar\footnote{
Note that the Three Musketeers, whose pulsar properties are very similar to those of J1741 (see Table~\ref{table:4PSRs2}), show neither FUV nor H$\alpha$ BSNe, indicating high ionization of the ISM around these pulsars. 
Moreover, their X-ray PWNe are drastically different from each other \citep{Posselt2017, Birzan2016, Posselt2015} 
and from the J1741 PWN. The differences can be associated with different geometries of their pulsar winds and/or different local ISM properties.}. 
In Table~\ref{table:BSNe} the properties of the J1741 FUV BSN are compared with those of the
the previously detected BSNe of the nearby millisecond pulsars J0437--4715 \citep{Rangelov2016} and J2124--3358 \citep{Rangelov2017}, which we denote J0437 and J2124 hereafter. Note that the spin-down powers $\dot{E}$ of these millisecond pulsars are only slightly lower than that of J1741, and all the three pulsars have been detected in the GeV $\gamma$-ray range.

At first glance, the closed-shell shape of the J1741 BSN looks drastically different from those of the two other BSNe. The J0437 BSN looks like a classic bow shock, whose shape is excellently described by 
a simple equation 
\citep{Wilkin1996}.
The forward part of the J2124 BSN 
resembles the bow shock of J0437, but the brightness distribution is asymmetric with respect to the proper motion direction, and the bow shock narrows behind the pulsar, apparently forming a bubble with a hint of a Guitar-like structure \citep{Brownsberger2014}. The difference of the shapes of the BSN images may be associated with  
different geometries of the pulsar wind outflow, different local ISM pressures and/or 
magnetic fields, 
and different inclinations of the pulsar's velocity to the line of sight (possibly the smallest for J1741).

\citet{Romani2010} suggested that the transverse elongation of the J1741 BSN is due to 
the pulsar wind being concentrated in the equatorial plane of the rotating pulsar whose spin axis is aligned with the pulsar's velocity. 
However, if 
our assumption that only the bulge of the J1741 BSN is the actual forward bow shock
is correct,
then the apparent difference with the two other FUV/H$\alpha$ BSNe 
could be explained, at least partly, 
by the projection effect: if the velocity vector is
close to the line of sight, one will naturally get an observed shape, 
irrespective to the spin-velocity alignment.  
This would imply that the actual pulsar's velocity can significantly exceed its transverse velocity, $v_\perp \approx 140 d_{270}$ km s$^{-1}$, 
and the actual lengths of structures aligned with the velocity vector, such as the $\sim 10''$ ($\simeq 4.0\times 10^{16}$ cm in the sky projection) length of the compact PWN component,  are significantly longer.  Deeper FUV imaging would be most useful to examine this interpretation and estimate the velocity and the lengths more accurately.

We see from Table~\ref{table:BSNe} that the J1741 BSN has the smallest projected distance between the pulsar and the forward bow shock apex,  $l_{a,\perp} = 4.4\times 10^{15} d_{270}$ cm, 
which is factors of 5.2 and 3.4 smaller than for the J0437 and J2124 BSNe, respectively.
For an isotropic pulsar wind, the distance $l_a$ to the bow shock front apex measured along the velocity vector 
slightly exceeds the standoff distance, where the ram pressures of the ISM and the pulsar wind balance each other (e.g., \citealt{Brownsberger2014,Reynolds2017}): 
$l_a \approx 1.3 \left(\xi_w\dot{E}/4\pi c \rho v^2\right)^{1/2}$,
where 
$\xi_w$ is the fraction of $\dot{E}$ that powers the pulsar wind, $\rho$ is the mass density of the ambient ISM, $v=v_\perp/\sin i$ is the pulsar velocity, and $i$ is the velocity inclination with respect to the line of sight. The 
larger J1741's projected velocity together with a smaller inclination $i$ and/or a higher ISM density could explain 
the observed ratios of distances $l_{a,\perp}$, but an anisotropic pulsar wind could be another factor.

The FUV BSN of J1741 
looks 
brighter than those of J0437 and J2124. For instance, the (distance-independent) mean surface brightness  in the F125LP 
image of the 
Front region of J1741, ${\cal B}_s = 
110\pm 2$ cnts/(ks\,arcsec$^2$), 
is 
2.7 and 11 times higher than those of apex regions 
of J0437 and J2124 (the only regions for which ${\cal B}_s$ was measured in those BSNe). The brightness ratios would become a factor of $\sim 4$ 
larger if we account for the different ISM extinctions towards the pulsars (see Table~\ref{table:BSNe}). The high brightness of the J1741 FUV BSN makes it particularly suitable for FUV spectroscopy.

It is more difficult to compare the FUV count rates from entire BSNe because 
only parts of the J0437 and J2124 FUV BSNe were imaged in FUV.
However, assuming that the FUV brightness distribution is similar to that in H$\alpha$, we crudely estimated that if the total count rate of the entire J0437 FUV BSN were measured, it would be a factor of about 2 higher than the count rate for the entire J1741 BSN (e.g., $\approx$13 cnts/s in F125LP -- see Table~\ref{table:bsn_photometry}). 
Similar estimates for the faint J2124 BSN are even more uncertain, but most likely its total FUV count rate would be somewhat lower than that of the J1741 BSN.

It would also be interesting to compare the FUV energy fluxes and luminosities of the three FUV BSNe. Accurate estimates for these quantities require knowledge of their FUV spectra, but all the spectral information is contained in the count rates in only two filters, 
F125LP and F140LP, 
with the 
passband of the latter being within the passband of the former.
These count rates and, particularly, their ratio $\eta$ (see Tables~\ref{table:C125140} and \ref{table:BSNe})
can be used to 
estimate parameters of an assumed spectral model, as well as the FUV fluxes and luminosities
\citep{Rangelov2016}.
However, the fluxes and luminosities may be significantly different for spectra dominated by emission lines (e.g., from atoms or ions excited in passage of the bow shock) or a continuum (e.g.,
the continuum component of thermal emission from the shocked plasma or synchrotron radiation from electrons accelerated between the forward bow shock and the PWN termination shock, as suggested by \citealt{Bykov2017}). 


Assuming a continuum PL spectrum, in Section \ref{BSN} we estimated the PL slope $\alpha = -1.0 \pm 0.7$ and the unabsorbed FUV (1250--1850\,\AA, or 6.70--9.92 eV) energy flux 
$F_{\rm FUV,-14} \approx 54$ from the entire J1741 BSN.
This corresponds to the observed flux 
$F_{\rm FUV,-14}^{\rm obs} \approx 10$. 
Since $\eta = 1.6 \pm 0.1$ is the same for J0437 and J1741 BSNe, we can assume the same spectral slopes and 
estimate that the absorbed energy flux from the entire J0437 BSN is $\sim$2 times higher
than from the J1741 BSN. A similar 
flux ratio can be obtained if we scale the FUV flux from the J0437 BSN apex region, reported by \citealt{Rangelov2016}, and 
correct the error in the previous SBC flux calibration, reported by \citealt{Avila2019}, which resulted in $\sim$27\% lower flux for a given count rate in the F125LP filter. 
Assuming $E(B - V) = 0.22$ 
for J1741, and $E(B - V) = 0.05$ for J0437, the full J1741 BSN luminosity, $L_{\rm FUV} \approx 4.7\times 10^{30} d_{270}^2$ erg s$^{-1}$, is a factor of 5--7 
higher than the full J0437 FUV BSN luminosity. 
Estimates of the full J2124 nebula observed flux and luminosity are, again, very uncertain, but likely both the total observed flux 
and the total FUV luminosity are lower than those of the entire J1741 BSN.
The highest FUV luminosity of the J1741 BSN can be partly associated with the highest spin-down power 
of the J1741 pulsar and/or a higher ambient ISM density, but the dependence on these and other parameters depends on the BSN emission model.\\

Given the virtually coinciding images of the FUV and H$\alpha$ BSNe, the hypothesis of
the FUV BSN spectra being dominated by emission lines 
may look more plausible. 
Unfortunately, 
we can only guess which lines give a significant contribution. Obviously, it could only be strong lines of abundant elements. Since there are no 
H\,I or He\,I lines in this FUV range, one could consider,
for instance, the Balmer-alpha line of He\,II, at 
$\approx$1640 \AA\ (7.56 eV) or
the strong C\,IV 
resonance line with fine structure components at 1548.19 and 1550.77 \AA\ (around 8.00 eV). It is, however, difficult to explain lines of ionized species, particularly the highly ionized Carbon.
Since the BSNe are seen in H$\alpha$, the ISM should be mostly neutral ahead of them. It means that 
the initially neutral atoms, which can easier penetrate through the shock, should be ionized after passing through the shock, before 
their ions are excited to produce the lines.
This may require too high temperatures 
behind the forward shock and may lead to a FUV BSN brightness distribution 
%
different from 
the H$\alpha$ BSN brightness distribution, which has not been detected so far. On the other hand, \citet{Romani2010} report a few (forbidden) lines of once-ionized Nitrogen, Oxygen and Sulfur at optical wavelengths, so one can assume that lines of ionized elements can also be seen in the FUV J1741 BSN.  
To resolve this problem, BSN FUV spectroscopy is required, supported by realistic BSN models. 

Another option could be FUV emission of molecular Hydrogen, possibly due to Ly$\alpha$ fluorescence (see, for example, \citealt{Wood2001,France2012} and \citealt{Martin2007}). However, 
the probability that three of the few pulsars observed in FUV are in molecular clouds seems to be very low.

One could also consider continuum emission due to the two-photon $2s \rightarrow 1s$ transition in Hydrogen 
atoms\footnote{We thank Marten van Kerkwijk for reminding us about this possibility.}.   
Using an analytical approximation for the two-photon emission spectrum \citep{Nussbaumer_Schmutz1984} and the F125LP throughput, we found that in order to explain the observed count rate 
(e.g., for the Half region of the BSN) by two-photon emission, 
the ratio of the 
number of photons emitted by this mechanism to the number of H$\alpha$ photons should be about 11--12, assuming $E(B-V)=0.22$. Such a ratio is significantly higher than $\lesssim2$ predicted by \citet{Kulkarni_Shull2023} for 
low-velocity ($v_{s}\lesssim 70$ km s$^{-1}$) shocks. This ratio should be even lower for the high-velocity J1741 BSN. Therefore, it 
is very unlikely that two-photon decay is the main source of the FUV emission.

At the wavelengths of the mentioned 
He\,II and C\,IV lines the F125LP and F140LP throughputs are very close to each other (see Figure~\ref{fig:filters}). Therefore, there should be at least one other line (or/and a continuum) in the 
$\approx$1230--1360 \AA\ (9.12--10.09 eV) range in order to be consistent with the count rate ratio $\eta > 1$. Plausible candidates are the line complexes of OI lines at 1302--1306 \AA\ (9.49--9.52 eV), C\,I lines around 1277 \AA\ (9.72 eV), and N\,I lines around 1243 \AA\ (9.98 eV). 
Only FUV spectroscopy can 
convincingly show which (if any) of these lines 
are present in the BSN emission.

As we still do not know the wavelength(s) and strength(s) of the putative lines,
the flux and luminosity estimates are highly uncertain in a line-dominated FUV BSN spectrum.
As we discussed in 
Section \ref{BSN}, we can only estimate conservative lower limits on the observed FUV flux,
$F_{\rm FUV,-14}^{\rm obs} > 5.8$, 
and luminosity, $L_{\rm FUV} > 2.9 \times 10^{30} d_{270}^{2}$ erg s$^{-1}$.
The lower limits on the line-dominated FUV flux 
translate into the lower limit of 2.5 on the observed (absorbed) FUV-to-H$\alpha$ flux ratio 
while the ratio of the unabsorbed fluxes (and luminositie) is $>8$. These limits are about 1.6--1.8 lower than the flux ratios, 4.1 and 14 for the observed and unabsorbed fluxes, respectively, that follow from the above-estimated fluxes for the PL spectrum model.

Similar excesses of FUV flux over H$\alpha$ flux were obtained for the two other BSNe known \citep{Rangelov2016,Rangelov2017}. For instance, for the PL model of the J0437 BSN, 
the absorbed (unabsorbed) FUV flux 
is a factor of 5--7 (6--9) 
higher than the H$\alpha$ flux. 
For the J2124 BSN, 
the ratio of unabsorbed fluxes, $F_{\rm FUV}/F_{{\rm H}\alpha} \approx 3$, 
for a 
flat $f_{\rm \lambda}$ spectrum, 
but 
a factor of up to 2 higher values are not excluded.
If the FUV spectra of the J0437 and J2124 BSNe are line-dominated, the
$F_{\rm FUV}/F_{\rm H \alpha}$ ratios could be lower,
but they remain $>1$.

Since the photon energies in the FUV range, $\sim$(6.7--10) eV, exceed the energy 1.889 eV of H$\alpha$ photons by a factor of about  3.5--5.3, the ``photon luminosities'' of the FUV and H$\alpha$ BSNe are very close to each other. It is difficult to reconcile this finding with the line-dominated model of the FUV BSN spectra
because it would require extremely high abundances of chemical elements responsible for the lines. 
We believe that most likely the FUV BSN emission is produced in a hot shocked plasma flow behind the bow shock, 
whose spectrum contains both a continuum component and spectral lines.
Relative contributions of these component can hardly be inferred without modeling and further observations.
\\

Thus, we can summarize our main findings on the J1741 and the two other FUV BSNe as follows. 
\begin{itemize}
 \item   The three detected FUV BSNe (as well as their Balmer-line counterparts) are created by pulsars with very different periods and ages but with close spindown powers, $\dot{E} = (5.5$--$9.5)\times 10^{33}$ erg s$^{-1}$, and transverse velocities $v_\perp$ in the range of 100--150 km s$^{-1}$.
 All the three pulsars were detected in GeV $\gamma$-rays.
 \item There are many pulsars with similar $\dot{E}$ and $v_\perp$ that do not show BSNe, neither in FUV nor in H$\alpha$. It means that these parameters alone are not responsible for the presence of a BSN. Apparently, local ISM properties, such as density, pressure, and degree of ionization,
 are more important for BSN formation.
 \item No FUV BSNe have been detected without Balmer-line counterparts. This 
 implies presence of 
 neutral Hydrogen atoms (cold ISM) ahead of the forward shock, which get heated and excited in the shock passage. 
 The residual neutral Hydrogen behind the shock front cannot be responsible for the observed FUV emission, but emission from a partly ionized plasma (e.g., bremsstrahlung) can provide a significant contribution. 
 \item  All the FUV BSNe are spatially coincident with their Balmer-line counterparts, and the brightness distributions within the Balmer-line and FUV BSNe look similar. It suggests that they are created by similar processes, such as emission from a partly ionized plasma behind the forward shock. 
 \item We are not aware of Balmer-line BSNe without FUV counterparts, but six of the nine known H$\alpha$ BSNe have not been imaged in FUV.
 \item The FUV BSN luminosities, $L_{\rm FUV}\sim$(1--$5)\times 10^{30}$ erg s$^{-1}$, possibly correlate with $\dot{E}$. They exceed the H$\alpha$ luminosities 
 by a factor of 
 3--14 for all the three BSNe. It means that the FUV emission is not dominated by spectral lines because it would require unrealistically high abundances of elements heavier than Hydrogen.  
\item We presume that the FUV BSN emission comes from the ISM heated and ionized while passing through the forward bow shock. Its FUV spectrum likely includes both the continuum component and emission lines of abundant elements that can be studied in deep spectroscopic observations.
\end{itemize}

To conclude, the nature of the FUV BSNe remains very puzzling. 
To understand it, spectroscopic observations of the known FUV BSNe and a search for new FUV BSNe, supplemented by realistic models of these objects, is needed.

\begin{acknowledgements}
    Support for program \#17155 was provided by NASA through a grant HST-GO-17155 from the Space Telescope Science Institute, which is operated by the Association of Universities for Research in Astronomy, Inc., under NASA contract NAS 5-26555. The work of V.\,A. was supported by the Russian Science Foundation grant 24-12-00320, \url{https://rscf.ru/project/24-12-00320/}. 
    The work of Y.\,S.  was partially supported by the baseline project FFUG-2024-0002 of the Ioffe Institute.
    We are grateful to Roger Romani for useful suggestions.
    We thank A.\,V. Karpova and D.\,A. Zyuzin for providing the reduced {\em Chandra} data.
    We express our thanks to the referee Marten van Kerkwijk for careful reading of the manuscript and thoughtful remarks and suggestions.
    This work has made use of data from the European Space Agency (ESA) mission
    {\it Gaia} (\url{https://www.cosmos.esa.int/gaia}), processed by the {\it Gaia}
    Data Processing and Analysis Consortium (DPAC,
    \url{https://www.cosmos.esa.int/web/gaia/dpac/consortium}). Funding for the DPAC
    has been provided by national institutions, in particular the institutions
    participating in the {\it Gaia} Multilateral Agreement. This research has made use of data obtained from the {\it Chandra} Data Archive provided by the {\it Chandra} X-ray Center (CXC). 
    This research made use of the following software: IRAF 
    \citep{Tody1986,Tody1993}, XSPEC \citep{Arnaud1996}, DrizzlePac (\url{https://www.stsci.edu/scientific-community/software/drizzlepac}), EsoRex \citep{2015ESO}, and Python packages Matplotlib \citep{Hunter:2007}, emcee \citep{Foreman-Mackey2013}, NumPy \citep{harris2020array}, Astropy \cite{AstropyCollaboration2022}
\end{acknowledgements}



%
%

\bibliography{aanda_rev_clean_4}{}
\bibliographystyle{aa}

\end{document}